# The Response Matrix Discrete Ordinates Solution to the 1D Radiative Transfer Equation

Barry D. Ganapol
Department of Aerospace and Mechanical Engineering
University of Arizona
Ganapol@cowboy.ame.arizona.edu

**ABSTRACT**
The discrete ordinates method (DOM) of solution to the 1D radiative transfer equation has been an effective method of solution for nearly 70 years. During that time, the method has experienced numerous improvements as numerical and computational techniques have become more powerful and efficient. Here, we again consider the analytical solution to the discrete radiative transfer equation in a homogeneous medium by proposing a new, and consistent, form of solution that improves upon previous forms. Aided by a Wynn-epsilon convergence acceleration, its numerical evaluation can achieve extreme accuracy as demonstrated by comparison with published benchmarks. Finally, we readily extend the solution to a heterogeneous medium through the star product formulation producing a novel benchmark for closed form Henyey-Greenstein scattering as an example.

## I. INTRODUCTION

Every few years or so, someone comes up with a new numerical or analytical scheme to solve the radiative transfer equation (RTE) in 1D plane parallel geometry. This presentation is another one of those. The coherent scattering radiative transfer equation (RTE), best known for its application in investigations of solar radiation in planetary atmospheres, the Earth's atmosphere and oceans, and in the transport of neutrons in scattering, absorbing and fissioning media, is the focus. Its solution, in terms of discrete streams, gained prominence when initially suggested by Wick [25] and subsequently applied to general anisotropic scattering media with polarization by Chandrasekhar [2,3]. In the years following, many investigators contributed to the improvement of what we now know as the discrete-ordinates (numerical) method (DOM) of solution. Of particular note are the theoretical and applied contributions of Carlson, Dave, Devaux, Garcia, Gelbard, Irvine, Lenoble, Liou, Nakajima, Samuelson, Segatto, Siewert, Stamnes, Wiscombe and Yamamoto [1,4,5,7,8,11,13,14,15,17,18,20,21,26,28] (to name a few) all of whom influenced the form of the solution that exists today. Arguably, two individuals, K. Stamnes, C. E. Siewert along with R. M. Garcia particularly stand out for their contributions.

These researchers have, each in their own way and largely independently, made significant contributions to the modern day numerical implementation of DOM. Specifically, Stamnes and his co-authors [21,22] changed the way we find the eigenvalues and eigenvectors of the solution to the ODEs resulting from angular discretization. They relegated this essential aspect of the calculation to improved numerical linear algebra software demonstrating significant improvement in accuracy and reduction in computational effort. In general, therefore, largely because of Stamnes' effort, we can now consider the eigenvalues and eigenvectors as known, possibly with the exception of cases of extremely forward peaked scattering requiring exceptionally high order angular quadrature. On the other hand, Siewert, with Garcia, seeking exponential solutions embraced the linear algebra form of solution, providing a concise theoretical/numerical analysis called the Analytical Discrete Ordinates (ADO) method. If one had to make a distinction between the approaches of the two, it would be practicality versus accuracy. Stamnes is primarily interested in the application of the theory to actual atmospheric radiative transfer, where accuracy is generally four significant digits; whereas, Siewert is concerned with generating high accuracy results on the order of five or more significant digits to serve as benchmarks. Without a doubt, the ADO method has generated the highest quality numerical solutions to some of the most comprehensive transport problems to date—and not just for radiative transfer. Essentially, through their many publications, discrete ordinate methods have advanced into the modern computational era.

So-- one might therefore ask-- "What else can be done to improve the DOM or ADO methods?" As will soon become evident, we answer this question with a new more efficient form of solution for a homogeneous participating medium, an improved post processor for the angular intensity and an analytical response matrix solution for heterogeneous media. However, to properly make the case for a new solution, we first need to examine the current DOM.

**I.1 The Discrete Ordinate Equations**

One can view the derivation of the method of discrete ordinates in several different ways. The first is to consider a collocation of multiple particle streams, where a set of discrete values replaces the continuous directional variable $\mu$. Hence, the coherent 1D radiative transfer equation for photons moving in direction $\mu$ at optical depth $\tau$ in a homogeneous medium of scattering albedo $\omega$ and scattering phase function $f(\mu',\mu)$,

$$\left[\mu\frac{\partial}{\partial\tau}+1\right]I(\tau,\mu) = \omega\int_{-1}^{1}d\mu' f(\mu',\mu)I(\tau,\mu') + Q(\tau,\mu), \tag{1}$$

becomes

$$\left[\mu_m\frac{\partial}{\partial\tau}+1\right]I(\tau,\mu_m) = \omega\sum_{m'=1}^{2N}\omega_{m'} f(\mu_m,\mu_{m'})I(\tau,\mu_{m'}) + Q(\tau,\mu_m) + E_G(\tau,\mu_m;N). \tag{2a}$$

for direction $\mu_m$, $m=1,2,...,2N$. Thus, $\pm\mu_m$ replaces $\mu$ in the angular intensity $I(\tau,\mu)$ with the restriction

$$\mu = \begin{cases} -\mu_m \\ \mu_{N+m} \equiv \mu_m \end{cases}, \quad m=1,...,N, \quad \mu_m > 0; \tag{2b}$$

and a Gauss quadrature approximates the collision integral

$$\int_{-1}^{1}d\mu' f(\mu,\mu')I(\tau,\mu') = \sum_{m'=1}^{2N}\omega_{m'} f(\mu_{m'},\mu)I(\tau,\mu_{m'}) + E_G(\tau,\mu;N),$$

where $E_G(\tau,\mu;N)$ is the quadrature error. We will exclusively assume a double Gauss quadrature on the intervals [-1,0) and (0,1] with ordinates $\mu_m$ and weights $\omega_m$.

With the Lagrange interpolated intensity [24]

$$I(\tau,\mu) = \sum_{m=1}^{2N}l(\mu_m)I(\tau,\mu_m) + E_L(\tau,\mu;N),$$

introduced into the integral term in Eq(1) and $\mu$ set to $\mu_m$, the identical form of Eq(2a) results with interpolation error $E_L(\tau,\mu;N)$. $l(\mu_m)$ is the Lagrange interpolating polynomial. The two derivations are essentially equivalent except for the form of the error.

The discrete ordinates equation of radiative transfer results by ignoring the quadrature or interpolation errors

$$\left[\mu_m \frac{\partial}{\partial \tau}+1\right]I_m(\tau)=\omega\sum_{m'=1}^{2N}\alpha_{m'}f(\mu_{m'},\mu_m)I_{m'}(\tau)+Q(\tau,\mu_m), \qquad (3a)$$

for $m=1,2,...,2N$, where the true and approximate intensities are related by

$$I(\tau,\mu_m)=I_m(\tau)+\varepsilon(\tau,\mu_m).$$

There are several points to note. Specifically, for demonstration purposes, we primarily consider the azimuthally integrated RTE (zeroth order Fourier component) with a fixed source and possible sources at the slab surfaces to be included momentarily. When expressed in terms of spherical harmonics [7], the Fourier components decouple and one treats the $m^{th}$ component in exactly the same way as the zeroth component. To partially substantiate this claim, we do consider an example of a Fourier component other than the zeroth. In addition, the errors introduced by either the quadrature approximation or interpolation provide an analytical estimate of the actual DOM error, $\varepsilon(\tau,\mu_m)$ –– but we will not pursue this further. As an alternative, we address the actual DOM error through a Wynn-epsilon acceleration [19]. Finally, since the RTE becomes a set of 2N-coupled ODEs, the resultant intensity approximation should read $I_m(\tau;N)$ –– but we suppress the N-dependence until required.

At this point, one must decide on a specific method of solution for Eqs(3a). The radiative transfer and neutron transport communities differ in their choice. For neutrons, the common solution, called the $S_N$ (Segment N) method [1], discretizes the spatial coordinate as well. This is called diamond differencing since a diamond is the figure made when lines connect the centers of the sides of the rectangle made in ($\tau,\mu$) phase space. Starting with the boundary condition at the top surface, $\tau = 0$, of a homogeneous slab

$$I_m(0) = g_m, \quad m = N+1,...,2N, \tag{3b}$$

where $g(\mu)$ is a known distribution of entering particles, the solution tracks the particles in the positive direction to the lower boundary, $\tau = \tau_0$, initially assuming the intensities in the negative direction to be zero. Then, restarting from the condition at the bottom boundary

$$I_m(\tau_1) = h_m, \quad m = 1,...,N, \tag{3c}$$

one "sweeps" in the negative $\mu$- direction using the intensities just found in the positive direction in the collision term to estimate the scattered contribution. The iteration (sweeps) continues until convergence. The $S_N$ method has been the preferred solution for neutron transport as it does not suffer from the instability inherent in the spatially continuous form since eigenvalues and eigenvectors are not required. The method can be quite accurate for highly heterogeneous slabs under several hundred mean free paths. Most importantly, it extends to multi-dimensions. However, it is not appropriate for deep penetration, is not analytical and can require significant computer resources; nevertheless, it is one of the primary computational transport methods.

The preferred solution of the RT community is to maintain a continuous spatial variable and analytically solve the set of ODEs. This is certainly the most elegant solution and enables deep penetration. While seemingly the best way forward, the analytical solution has an inherent difficulty that requires careful consideration unlike the $S_N$ method. In particular, the solution in the form of exponentials of eigenvalues, which are real and occur in positive/negative pairs [12], is intrinsically unstable. The large disparity in the exponential solutions for positive and negative exponents causes the difficulty. Almost without exception, early proposers of exponential analytical forms discuss this issue. At one point, researchers thought that the DOM was not appropriate for slabs larger than an optical depth [17]. Researchers later found, as with many such instabilities, the way one expresses the solution enables the DOM. With an ad-hoc scaling [23], the solution becomes appropriate for slabs of virtually any mean free path. As will be shown, scaling, as a separate step in the solution, is indeed unnecessary.

In this presentation, we investigate a new expression of the solution. The new form brings with it a theoretically inline simplification previously not reported. Yamamoto [28], however, came close and Nakajima and King [16] even closer. The idea is straightforward and comes directly from the theory of solutions to second order ODEs as will now be explained.

## II. THEORY

To summarize, the equations of the DOM in a 1D homogeneous slab $(0 < \tau < \tau_0)$ are

$$\left[\mu_m \frac{\partial}{\partial \tau} + 1\right] I_m(\tau) = \omega \sum_{m'=1}^{2N} \omega_{m'} f(\mu_{m'}, \mu_m) I_{m'}(\tau) + Q(\tau, \mu_m) \qquad (4a)$$

$$I_m(0) = g_m, \ m = N+1, ..., 2N$$
$$I_m(\tau_0) = 0, \ m = 1, ..., N, \qquad (4b)$$

where we assume an anisotropic entering intensity on the top boundary and no incoming intensity on the bottom boundary. The fixed volume source is general. Our primary goal will be to develop a robust, efficient numerical algorithm giving extreme accuracy of at least 7-places for the determination of the angular intensity in any direction and optical depth. The scalar intensity and flux,

$$I(\tau) \equiv \int_{-1}^{1} d\mu I(\tau, \mu) \simeq \sum_{m=1}^{2N} \omega_m I_m(\tau) = \mathbf{1}^T \mathbf{W} \mathbf{I}(\tau)$$

$$q(\tau) \equiv \int_{-1}^{1} d\mu \mu I(\tau, \mu) \simeq \sum_{m=1}^{2N} \omega_m \mu_m I_m(\tau) = \boldsymbol{\mu}^T \mathbf{W} \mathbf{I}(\tau), \qquad (5a,b)$$

follow, in the $N$th approximation respectively, where

$$\mathbf{1}^T \equiv \begin{bmatrix} 1 & 1 & ... & 1 \end{bmatrix}$$

$$\boldsymbol{\mu}^T \equiv \begin{bmatrix} \mu_1 & \mu_2 & ... & \mu_{2N} \end{bmatrix}$$

and

$$\mathbf{I}(\tau) \equiv \begin{bmatrix} I_1(\tau) & I_2(\tau) & .... & I_{2N}(\tau) \end{bmatrix}^T.$$

Additionally, of particular importance in applications, are the slab surface reflectance and transmittance

$$R_f \equiv \gamma_f \int_{-1}^{0} d\mu \mu I(0,\mu) \simeq \gamma_f \sum_{m=1}^{2N} \omega_m \mu_m I_m(0)$$

$$T_n \equiv \gamma_n \int_{0}^{1} d\mu \mu I(\tau_1,\mu) \simeq \gamma_n \sum_{m=1}^{2N} \omega_m \mu_m I_m(\tau_0),$$

(5c,d)

where $\gamma_f$ and $\gamma_n$ are coefficients to be adjusted according to the various definitions of reflectance and transmittance found in the literature.

There are two major elements in achieving the goal of robust, efficient and accurate. The first is to establish a sound and stable theory of solution through linear algebra. The second is the numerical delivery, which seems to be where significant innovation takes place. While a stable algorithm is necessary, it does not guarantee accuracy and efficiency. Specifically, one must acknowledge that the result of the computation is nothing more than the $N^{th}$ order approximation and use this to advantage in order to achieve extreme accuracy.

### II.1. Reduction to a First Order ODE

Noting the differential/algebraic nature of Eqs(4), numerous attempts at an efficient numerical algorithm have appeared in the literature—the most recent of which involve numerical linear algebra. These methods center on the determination of eigenvalues (or zeros of a polynomial in the case of the LTSN method [18]) and a solution representation as the span of the complete set of eigensolutions. However, because of instability, caution is necessary since the form of solution leads to numerical difficulty [e.g., 8,17,18,22,28]. Therefore, we will derive an alternative form to overcome this difficulty as a natural part of the solution and not, as has been done previously [17,23], an afterthought.

The basis of the linear algebra approach is the reformulation of Eqs(4) in terms of positively and negatively directed photons to give the following vector equations in each direction:

$$\frac{d\mathbf{I}^{\pm}(\tau)}{d\tau} = \mp \mathbf{M}^{-1}\left(\mathbf{I}_N - \mathbf{C}^{\pm\pm}\right)\mathbf{I}^{\pm}(\tau) + \mathbf{M}^{-1}\mathbf{C}^{\pm\mp}\mathbf{I}^{\mp}(\tau) \pm \mathbf{M}^{-1}\mathbf{Q}^{\pm}(\tau), \qquad (6a)$$

where

$$\mathbf{M} \equiv diag\{\mu_m\}$$

$$\mathbf{I}^{\mp}(\tau) \equiv \left[ I_{\binom{1}{N+1}}(\tau) \quad I_{\binom{2}{N+2}}(\tau) \quad .... \quad I_{\binom{N}{2N}}(\tau) \right]^T$$

$$\mathbf{Q}^{\mp}(\tau) \equiv \left[ Q_{\binom{1}{N+1}} \quad Q_{\binom{2}{N+2}} \quad .... \quad Q_{\binom{N}{2N}} \right]^T \qquad (6b,c,d,e,f)$$

$$\mathbf{CW} \equiv \begin{bmatrix} \mathbf{C}^{--} & \mathbf{C}^{-+} \\ \mathbf{C}^{+-} & \mathbf{C}^{++} \end{bmatrix} \mathbf{W} = \{\omega_m f(\pm\mu_{m'}, \pm\mu_m); j, m = 1,...,N\}$$

$$\mathbf{W} \equiv diag\{\omega_m\}.$$

$\mathbf{I}_N$ is the identity matrix of order $N$ and that boldface indicates either a vector or matrix. The assumed phase function is a truncated (at $L$) Legendre expansion

$$f(\mu',\mu) \equiv \frac{\omega}{2}\sum_{l=0}^{L}\beta_l P_l(\mu') P_l(\mu). \qquad (7)$$

Our intent is to solve the set of Eqs(6a)

$$\begin{bmatrix} \dfrac{d\mathbf{I}^-(\tau)}{d\tau} \\ \dfrac{d\mathbf{I}^+(\tau)}{d\tau} \end{bmatrix} = \begin{bmatrix} \alpha & -\beta \\ \beta & -\alpha \end{bmatrix} \begin{bmatrix} \mathbf{I}^-(\tau) \\ \mathbf{I}^+(\tau) \end{bmatrix} + \mathbf{M}^{-1} \begin{bmatrix} \mathbf{Q}^-(\tau) \\ \mathbf{Q}^+(\tau) \end{bmatrix} \quad (8a)$$

with boundary conditions

$$\begin{aligned} \mathbf{I}^+(0) &= \mathbf{g} \\ \mathbf{I}^-(\tau_0) &= \mathbf{0}, \end{aligned} \quad (8b)$$

where

$$\begin{aligned} \alpha &\equiv \mathbf{M}^{-1}\left(\mathbf{I}_N - \mathbf{C}^{++}\mathbf{W}\right) \\ \beta &\equiv \mathbf{M}^{-1}\mathbf{C}^{+-}\mathbf{W}. \end{aligned} \quad (8c)$$

While Eqs(8) are solvable, a more convenient and informative solution follows.

## II.2. Reduction to a Second Order ODE

By following the approach in [21] to give a reduction in effort in the solution of Eqs(8), we form two new dependent variables

$$\psi^\pm(\tau) \equiv \mathbf{I}^+(\tau) \pm \mathbf{I}^-(\tau). \quad (9)$$

If

$$\begin{aligned} \tilde{A} &\equiv (\alpha + \beta)(\alpha - \beta) \\ \xi^\pm(\tau) &\equiv \mathbf{M}^{-1}\left(\mathbf{Q}^+(\tau) \pm \mathbf{Q}^-(\tau)\right) \\ q(\tau) &\equiv \dfrac{d\xi^-(\tau)}{d\tau} - (\alpha + \beta)\xi^+(\tau), \end{aligned} \quad (10a,b,c)$$

one arrives at the following auxiliary ODE for $\psi^+(\tau)$:

$$\frac{d^2\psi^+(\tau)}{d\tau^2} - \tilde{A}\psi^+(\tau) = q(\tau). \tag{11a}$$

In addition, we find for $\psi^-(\tau)$

$$\psi^-(\tau) = -(\alpha+\beta)^{-1}\left[\frac{d\psi^+(\tau)}{d\tau} + \xi^-(\tau)\right], \tag{11b}$$

once $\psi^+(\tau)$ has been determined. With $\psi^\pm(\tau)$ determined, the intensity vectors in the two directions become

$$I^\pm(\tau) = \frac{1}{2}\left[\psi^+(\tau) \pm \psi^-(\tau)\right]. \tag{12}$$

### II.3. The General Solution

According to theory, the solution for $\psi^+(\tau)$ is a sum of solutions to the homogeneous and particular equations

$$\psi^+(\tau) = \psi_h^+(\tau) + \psi_P^+(\tau) \tag{13}$$

The next step is to diagonalize the Jacobian matrix $\tilde{A}$ [9]

$$\tilde{A} = T\lambda^2 T^{-1}, \tag{14a}$$

where the columns of $T$ are the eigenvectors of $\tilde{A}$ possessing $N$ eigenvalues

$$\lambda^2 \equiv diag\{\lambda_k^2\}, \tag{14b}$$

which, because of the symmetry of rotationally invariant scattering,

$$C^{++} = C^{--}$$
$$C^{+-} = C^{-+},$$

are real, positive and conveniently expressed as squares. $\boldsymbol{T}$ and $\lambda_k^2$ are the reduced eigenvectors and eigenvalues [21]. Of course, we assume that the set of $N$ eigenvectors of $\tilde{\boldsymbol{A}}$ are complete; otherwise, diagonalization is not possible. There is nothing new in the above decomposition as it is well known, but what follows is not.

**II.4. Homogeneous and Particular Solutions**
When one defines

$$\boldsymbol{\Theta}(\tau) \equiv \boldsymbol{T}^{-1}\boldsymbol{\psi}^{+}(\tau), \tag{15a}$$

and introduces the matrix decomposition into Eq(11a) with the diagonalization, there results

$$\left[\boldsymbol{I}\frac{d^2}{d\tau^2} - \boldsymbol{\lambda}^2\right]\boldsymbol{\Theta}(\tau) = \tilde{\boldsymbol{q}}(\tau) \tag{15b}$$

with

$$\tilde{\boldsymbol{q}}(\tau) \equiv \boldsymbol{T}^{-1}\boldsymbol{q}(\tau). \tag{15c}$$

We therefore seek solutions to the diagonalized homogenous and particular ODE forms

$$\left[\boldsymbol{I}\frac{d^2}{d\tau^2} - \boldsymbol{\lambda}^2\right]\boldsymbol{\Theta}_h(\tau) = \boldsymbol{0}$$

$$\left[\boldsymbol{I}\frac{d^2}{d\tau^2} - \boldsymbol{\lambda}^2\right]\boldsymbol{\Theta}_P(\tau) = \tilde{\boldsymbol{q}}(\tau), \tag{16a,b}$$

where

$$\Theta_h(\tau) = T^{-1}\psi_h^+(\tau)$$
$$\Theta_P(\tau) = T^{-1}\psi_P^+(\tau);$$
(16c)

and from Eq(13)

$$\psi^+(\tau) = T\Theta_h(\tau) + T\Theta_P(\tau).$$
(17)

Thus, diagonalization expresses the solution as independent modes each of which satisfies

$$\left[\frac{d^2}{d\tau^2} - \lambda_k^2\right]\Theta_{hk}(\tau) = 0$$

$$\left[\frac{d^2}{d\tau^2} - \lambda_k^2\right]\Theta_{Pk}(\tau) = \tilde{q}_k(\tau).$$
(18a,b)

We now focus on finding two independent solutions to the homogeneous equation, Eq(18a) realizing that there are choices other than ordinary exponentials. In particular, the following homogeneous solutions, or combinations thereof, are permissible:

$$e^{-\lambda_k\tau}, e^{\lambda_k\tau}$$

$$e^{-\lambda_k\tau}, e^{-\lambda_k(\tau_0-\tau)}$$

$$sinh(\lambda_k\tau),\ cosh(\lambda_k\tau).$$

The first set leads to instability. The second set eliminates the instability [18,20,22] but are not linearly independent for a zero eigenvalue. No one has yet to suggest the last pair until now. Therefore, we assume the following set of linearly independent solutions:

$$\frac{sinh(\lambda_k\tau)}{sinh(\lambda_k\tau_0)}, \frac{sinh(\lambda_k(\tau_0-\tau))}{sinh(\lambda_k\tau_0)}$$
(19)

and readily see the first advantage— the homogeneous solution remains bounded for all $\tau$, $\tau_0$ and $\lambda_k$ and is therefore no longer in danger of becoming unstable. The general complementary solution for $\Theta_{hk}(\tau)$ is

$$\Theta_{hk}(\tau) = h(\lambda_k \tau)\Theta_{hk}(\tau_0) + h(\lambda_k(\tau_0 - \tau))\Theta_{hk}(0), \tag{20a}$$

where

$$h(\lambda_k \tau) \equiv \frac{sinh(\lambda_k \tau)}{sinh(\lambda_k \tau_0)}. \tag{20b}$$

Note that the stated (unknown) coefficients, $\Theta_{hk}(0)$ and $\Theta_{hk}(\tau_0)$, while still arbitrary, are set to the homogeneous solution at the slab boundaries and Eq(20a) gives the appropriate identities as $h(\lambda_k \tau)$ cycles from 0 to 1 for $\tau = 0, \tau_0$ and correspondingly, $h(\lambda_k(\tau_0 - \tau))$ from 1 to 0.

The Wronskian for these solutions is

$$W(\lambda_k) = -\frac{\lambda_k}{sinh(\lambda_k \tau_0)} \tag{20c}$$

indicating that they are indeed independent, even for the case of a zero eigenvalue found for a conservative medium ($\omega$ = 1). The independent solutions for the conservative case are therefore distinct

$$\lim_{\lambda_k \to 0} h(\lambda_k \tau) \equiv \frac{\tau}{\tau_0}$$

$$\lim_{\lambda_k \to 0} h(\lambda_k(\tau_0 - \tau)) \equiv \frac{\tau_0 - \tau}{\tau_0}. \tag{21a,b}$$

Thus, another advantage of this choice of homogeneous solutions over any of the others is that the conservative case is included and requires no further attention as is required in the current theory.

With the homogeneous solutions known, it is now a relatively simple matter to establish the particular solution from the method of variation of parameters as

$$\Theta_{Pk}(\tau) = -W^{-1}(\lambda_k) \begin{bmatrix} h(\lambda_k \tau) \int_{\tau}^{\tau_0} d\tau' h(\lambda_k(\tau_0 - \tau')) \tilde{q}_k(\tau') + \\ + h(\lambda_k(\tau_0 - \tau)) \int_{0}^{\tau} d\tau' h(\lambda_k \tau') \tilde{q}_k(\tau') \end{bmatrix}. \qquad (22)$$

Note that for convenience, the particular solution has been set to zero at the slab boundaries since one also has a choice of its analytical form through the limits of integration. In the conservative case for $\lambda_k = 0$, this expression limits to

$$\Theta_{Pk}(\tau) = -\frac{1}{\tau_0}\left[\tau \int_{\tau}^{\tau_0} d\tau'(\tau_0 - \tau')\tilde{q}_k(\tau') + (\tau_0 - \tau)\int_{0}^{\tau} d\tau' \tau' \tilde{q}_k(\tau')\right].$$

Continuing--from Eq(17), $\boldsymbol{\psi}^+(\tau)$ is

$$\boldsymbol{\psi}^+(\tau) \equiv \boldsymbol{H}(\tau)\boldsymbol{\psi}_h^+(\tau_0) + \boldsymbol{H}(\tau_0 - \tau)\boldsymbol{\psi}_h^+(0) + \boldsymbol{\psi}_P^+(\tau), \qquad (23a)$$

with

$$\boldsymbol{\psi}_P^+(\tau) = \boldsymbol{T}\boldsymbol{\Theta}_P(\tau), \qquad (23b)$$

and the matrix function $\boldsymbol{H}$

$$\boldsymbol{H}(\tau) \equiv \boldsymbol{T}\left[diag\left\{\frac{\sinh(\lambda_k \tau)}{\sinh(\lambda_k \tau_0)}\right\}\right]\boldsymbol{T}^{-1}. \qquad (23c)$$

In a higher level explicit form, $\boldsymbol{H}(\tau)$ is

$$\boldsymbol{H}(\tau) \equiv \left[\boldsymbol{T}\left[diag\{sinh(\lambda_k \tau_0)\}\right]\boldsymbol{T}^{-1}\right]^{-1}\left[\boldsymbol{T}\left[diag\{sinh(\lambda_k \tau)\}\right]\boldsymbol{T}^{-1}\right] \quad (24a)$$
$$= sinh^{-1}\left(\sqrt{\tilde{\boldsymbol{A}}}\tau_0\right)sinh\left(\sqrt{\tilde{\boldsymbol{A}}}\tau\right) = sinh\left(\sqrt{\tilde{\boldsymbol{A}}}\tau\right)sinh^{-1}\left(\sqrt{\tilde{\boldsymbol{A}}}\tau_0\right),$$

where

$$sinh\left(\sqrt{\tilde{\boldsymbol{A}}}\tau\right) \equiv \boldsymbol{T}\left[diag\{sinh(\lambda_k \tau)\}\right]\boldsymbol{T}^{-1} \quad (24b)$$

and

$$\sqrt{\tilde{\boldsymbol{A}}} \equiv \boldsymbol{T} diag\{\lambda_k\}\boldsymbol{T}^{-1}. \quad (24c)$$

This level of abstraction is suitable for **MATLAB**[TM] or **MAPLE**[TM] implementation. Also, the solution degenerates into the scalar solution in the case of a two directional rod, which is the solution to the diffusion equation.

From the vector elements of Eq(22), the particular solution vector becomes

$$\boldsymbol{\psi}_P^+(\tau) = -\boldsymbol{W}^{-1}\left[\begin{array}{c} \boldsymbol{H}(\tau)\int_\tau^{\tau_0} d\tau' \boldsymbol{H}(\tau_0 - \tau')\boldsymbol{q}(\tau') + \\ +\boldsymbol{H}((\tau_0 - \tau))\int_0^\tau d\tau' \boldsymbol{H}(\tau')\boldsymbol{q}(\tau') \end{array}\right] \quad (25a)$$

where the inverse of the Wronskian from Eq(20c) is

$$\boldsymbol{W}^{-1} \equiv -\boldsymbol{T}\left[diag\left\{\frac{sinh(\lambda_k \tau_0)}{\lambda_k}\right\}\right]\boldsymbol{T}^{-1}. \quad (25b)$$

Since

$$\psi_h^+(\tau) = \psi^+(\tau) - \psi_P^+(\tau),$$

Eq(23a) becomes

$$\psi^+(\tau) \equiv H(\tau)\left[\psi^+(\tau_0) - \psi_P^+(\tau_0)\right] + H(\tau_0 - \tau)\left[\psi^+(0) - \psi_P^+(0)\right] + \psi_P^+(\tau) \quad (26a)$$

and since the particular solution vanishes at the slab surfaces

$$\psi^+(\tau) \equiv H(\tau)\psi^+(\tau_0) + H(\tau_0 - \tau)\psi^+(0) + \psi_P^+(\tau). \quad (26b)$$

To complete the analysis, Eq(11b) gives

$$\psi^-(\tau) = -(\alpha + \beta)^{-1}\left[\frac{dH(\tau)}{d\tau}\psi^+(\tau_0) + \frac{dH(\tau_0 - \tau)}{d\tau}\psi^+(0) + \frac{d\psi_P^+(\tau)}{d\tau} + \xi^-(\tau)\right]. \quad (26c)$$

The solution to Eq(11a) expressed in **MATLAB**™ format is therefore

$$\psi^+(\tau) = \sinh^{-1}\left(\sqrt{\tilde{A}}\tau_0\right)\begin{bmatrix} \sinh\left(\sqrt{\tilde{A}}\tau\right)\psi^+(\tau_0) + \sinh\left(\sqrt{\tilde{A}}(\tau_0 - \tau)\right)\psi^+(0) - \\ -\tilde{A}^{-1/2}\int_\tau^{\tau_0} d\tau' \sinh\left(\sqrt{\tilde{A}}\tau\right)\sinh\left(\sqrt{\tilde{A}}(\tau_0 - \tau')\right)q(\tau') - \\ -\tilde{A}^{-1/2}\int_0^\tau d\tau' \sinh\left(\sqrt{\tilde{A}}(\tau_0 - \tau)\right)\sinh\left(\sqrt{\tilde{A}}\tau'\right)q(\tau') \end{bmatrix}$$

(27a)

and

$$\psi^-(\tau) = -(\alpha+\beta)^{-1} \sinh^{-1}\left(\sqrt{\tilde{A}}\tau_0\right) \cdot$$

$$\cdot \begin{bmatrix} \sqrt{\tilde{A}}\cosh\left(\sqrt{\tilde{A}}\tau\right)\psi^+(\tau_0) - \sqrt{\tilde{A}}\cosh\left(\sqrt{\tilde{A}}(\tau_0-\tau)\right)\psi^+(0) - \\ -\int_\tau^{\tau_0} d\tau' \cosh\left(\sqrt{\tilde{A}}\tau\right)\sinh\left(\sqrt{\tilde{A}}(\tau_0-\tau')\right)q(\tau') + \\ +\int_0^\tau d\tau' \cosh\left(\sqrt{\tilde{A}}(\tau_0-\tau)\right)\sinh\left(\sqrt{\tilde{A}}\tau'\right)q(\tau') \end{bmatrix} - (\alpha+\beta)^{-1}\xi^-(\tau).$$

(27b)

Note that all matrix functions commute.

At this time, $\psi^+(0)$ and $\psi^+(\tau_0)$ are still unknown. We now find these quantities as an intermediate step in the determination of the outgoing boundary intensities $I^-(0)$ and $I^+(\tau_0)$.

### II.5. Determination of Outgoing Intensities $I^-(0)$ and $I^+(\tau_0)$

No useful boundary information comes from Eq(26a) since it collapses to an identity at the boundaries. All the boundary information will therefore come from Eq(26b), which gives for $\tau = 0, \tau_0$

$$(\alpha+\beta)\psi^-(0) = -\left[\frac{dH(\tau)}{d\tau}\bigg|_{\tau=0}\psi^+(\tau_0) + \frac{dH(\tau_0-\tau)}{d\tau}\bigg|_{\tau=0}\psi^+(0) + v(0)\right]$$

$$(\alpha+\beta)\psi^-(\tau_0) = -\left[\frac{dH(\tau)}{d\tau}\bigg|_{\tau=\tau_0}\psi^+(\tau_0) + \frac{dH(\tau_0-\tau)}{d\tau}\bigg|_{\tau=\tau_0}\psi^+(0) + v(\tau_0)\right],$$

(28a)

where

$$v(\tau) \equiv \frac{d\psi_P^+(\tau)}{d\tau} + \xi^-(\tau). \tag{28b}$$

With the following definitions:

$$A \equiv \left.\frac{dH(\tau_0 - \tau)}{d\tau}\right|_{\tau=0} = T\left[diag\left\{-\frac{\lambda_k \cosh(\lambda_k \tau_0)}{\sinh(\lambda_k \tau_0)}\right\}\right]T^{-1}$$
$$= -\sqrt{\tilde{A}}\, \coth\left(\sqrt{\tilde{A}}\tau_0\right) \qquad (29a,b)$$

$$B \equiv \left.\frac{dH(\tau)}{d\tau}\right|_{\tau=0} = T\left[diag\left\{\frac{\lambda_k}{\sinh(\lambda_k \tau_0)}\right\}\right]T^{-1} = \sqrt{\tilde{A}}\, \csc\left(\sqrt{\tilde{A}}\tau_0\right),$$

there results

$$\left.\frac{dH(\tau_0 - \tau)}{d\tau}\right|_{\tau=\tau_0} = -B$$
$$\left.\frac{dH(\tau)}{d\tau}\right|_{\tau=\tau_0} = -A \qquad (30a,b)$$

to give for Eq(28a)

$$\begin{bmatrix} B & B \\ x^+ & -x^- \end{bmatrix}\begin{bmatrix} I^-(\tau_0) \\ I^+(\tau_0) \end{bmatrix} = \begin{bmatrix} x^- & -x^+ \\ -B & -B \end{bmatrix}\begin{bmatrix} I^-(0) \\ I^+(0) \end{bmatrix} + \begin{bmatrix} v(0) \\ v(\tau_0) \end{bmatrix}, \qquad (31a)$$

with

$$x^\pm \equiv \alpha + \beta \pm A. \qquad (31b)$$

The most numerically useful form for Eq(31a) (reason to be discussed below) comes from the rearrangement

$$\begin{bmatrix} x^- & -B \\ -B & x^- \end{bmatrix}\begin{bmatrix} I^-(0) \\ I^+(\tau_0) \end{bmatrix} = \begin{bmatrix} B & x^+ \\ x^+ & B \end{bmatrix}\begin{bmatrix} I^-(\tau_0) \\ I^+(0) \end{bmatrix} + \begin{bmatrix} v(0) \\ v(\tau_0) \end{bmatrix} \qquad (32a)$$

to give the outgoing intensity at both surfaces in terms of the incoming at both surfaces. On inversion (assuming it exists), we obtain the exiting intensities in terms of the slab response matrix $\boldsymbol{R}$

$$\begin{bmatrix} \boldsymbol{I}^-(0) \\ \boldsymbol{I}^+(\tau_0) \end{bmatrix} = \boldsymbol{R} \begin{bmatrix} \boldsymbol{I}^-(\tau_0) \\ \boldsymbol{I}^+(0) \end{bmatrix} + \begin{bmatrix} \tilde{\boldsymbol{s}}_1 \\ \tilde{\boldsymbol{s}}_2 \end{bmatrix}, \qquad (32b)$$

where

$$\boldsymbol{R} \equiv \begin{bmatrix} \boldsymbol{x}^- & -\boldsymbol{B} \\ -\boldsymbol{B} & \boldsymbol{x}^- \end{bmatrix}^{-1} \begin{bmatrix} \boldsymbol{B} & \boldsymbol{x}^+ \\ \boldsymbol{x}^+ & \boldsymbol{B} \end{bmatrix} \qquad (32c)$$

and the source,

$$\begin{bmatrix} \tilde{\boldsymbol{s}}_1 \\ \tilde{\boldsymbol{s}}_2 \end{bmatrix} \equiv \begin{bmatrix} \boldsymbol{x}^- & -\boldsymbol{B} \\ -\boldsymbol{B} & \boldsymbol{x}^- \end{bmatrix}^{-1} \begin{bmatrix} \boldsymbol{v}(0) \\ \boldsymbol{v}(\tau_0) \end{bmatrix}, \qquad (32d)$$

in terms of the particular solution and its derivative.

One can reduce the order of the inversion in Eq(32c) from order $2N$ to two inversions of order $N$ since $\boldsymbol{R}$ is the multiplication of two partitioned symmetric matrices with identically partitioned diagonal elements

$$\boldsymbol{R} \equiv \begin{bmatrix} (\boldsymbol{x}^- - \boldsymbol{B})^{-1} + (\boldsymbol{x}^- + \boldsymbol{B})^{-1} & (\boldsymbol{x}^- - \boldsymbol{B})^{-1} - (\boldsymbol{x}^- + \boldsymbol{B})^{-1} \\ (\boldsymbol{x}^- - \boldsymbol{B})^{-1} - (\boldsymbol{x}^- + \boldsymbol{B})^{-1} & (\boldsymbol{x}^- - \boldsymbol{B})^{-1} + (\boldsymbol{x}^- + \boldsymbol{B})^{-1} \end{bmatrix} \begin{bmatrix} \boldsymbol{B} & \boldsymbol{x}^+ \\ \boldsymbol{x}^+ & \boldsymbol{B} \end{bmatrix}. \qquad (33a)$$

Hence, on matrix multiplication, we arrive at a familiar form

$$\boldsymbol{R} \equiv \begin{bmatrix} \boldsymbol{T}_n & \boldsymbol{R}_f \\ \boldsymbol{R}_f & \boldsymbol{T}_n \end{bmatrix} \qquad (33b)$$

with

$$\begin{Bmatrix} R_f \\ T_n \end{Bmatrix} = (x^- - B)^{-1}(x^+ + B) \pm (x^- + B)^{-1}(x^+ - B). \tag{33c}$$

$T_n$ and $R_f$ are the slab transmission reflection matrices respectively both of which contribute to the exiting intensity.

Finally, the source in Eq(32b) is now

$$\begin{bmatrix} S_1 \\ S_2 \end{bmatrix} \equiv \begin{bmatrix} (x^- - B)^{-1} + (x^- + B)^{-1} & (x^- - B)^{-1} - (x^- + B)^{-1} \\ (x^- - B)^{-1} - (x^- + B)^{-1} & (x^- - B)^{-1} + (x^- + B)^{-1} \end{bmatrix} \begin{bmatrix} \mathbf{v}(0) \\ \mathbf{v}(\tau_0) \end{bmatrix}. \tag{34}$$

Thus, once the slab response matrix $R$ is determined from knowledge of the phase function, slab albedo and slab thickness, the exiting intensities are also determined from the known entering and volume sources. The exiting distributions then give the interior distribution as will now be shown.

### II.6. The Angular Intensity

The general solution for the intensity vector at any $\tau$ follows from Eq(32b). Recall, the angular intensities in the forward and backward directions are

$$I^\pm(\tau) = \frac{1}{2}\left[\psi^+(\tau) \pm \psi^-(\tau)\right].$$

If

$$A^\mp(\tau) \equiv H(\tau) \mp (\alpha + \beta)^{-1} \frac{dH(\tau)}{d\tau}$$

$$= \sinh^{-1}\left(\sqrt{\tilde{A}}\tau_0\right) \begin{bmatrix} \sinh\left(\sqrt{\tilde{A}}\tau\right) \pm \\ \pm (\alpha + \beta)^{-1} \sqrt{\tilde{A}} \cosh\left(\sqrt{\tilde{A}}\tau\right) \end{bmatrix} \tag{35a}$$

$$S^{\pm}(\tau) \equiv \pm \left[ \psi_P^+(\tau) + \frac{d\psi_P^+(\tau)}{d\tau} + \xi^-(\tau) \right] = \pm S(\tau), \qquad (35b)$$

and by substitution of these expressions with Eqs(26) into Eq(12), there results

$$I^{\pm}(\tau) = \frac{1}{2} \left\{ \begin{array}{l} A^{\mp}(\tau)\left[I^+(\tau_0) + I^-(\tau_0)\right] + \\ + A^{\mp}(\tau - \tau_0)\left[I^+(0) + I^-(0)\right] \pm (\alpha + \beta)^{-1} S(\tau) \end{array} \right\}. \qquad (36)$$

Through rearrangement and in combination with Eq(32b), the last equation, expressed as the entire angular flux vector, is

$$\begin{bmatrix} I^-(\tau) \\ I^+(\tau) \end{bmatrix} = \frac{1}{2} \left\{ \begin{bmatrix} A^+(\tau - \tau_0) & A^+(\tau) \\ A^-(\tau - \tau_0) & A^-(\tau) \end{bmatrix} R + \begin{bmatrix} A^+(\tau) & A^+(\tau - \tau_0) \\ A^-(\tau) & A^-(\tau - \tau_0) \end{bmatrix} \begin{bmatrix} I^-(\tau_0) \\ I^+(0) \end{bmatrix} + \right.$$

$$\left. + \frac{1}{2} \left\{ \begin{bmatrix} A^+(\tau - \tau_0) & A^+(\tau) \\ A^-(\tau - \tau_0) & A^-(\tau) \end{bmatrix} \tilde{s} \pm \begin{bmatrix} (\alpha + \beta)^{-1} & 0 \\ 0 & (\alpha + \beta)^{-1} \end{bmatrix} \begin{bmatrix} S(\tau) \\ S(\tau) \end{bmatrix} \right\} \qquad (37)$$

to explicitly give the angular intensity at any optical depth for known incoming intensities and volume source. The approximate integrated quantities of Eqs(5), are readily found from the last expression via inner products

$$\begin{aligned} I(\tau) &= \mathbf{1}^T \mathbf{W} \mathbf{I}(\tau) \\ q(\tau) &= \boldsymbol{\mu}^T \mathbf{W} \mathbf{I}(\tau) \\ R_f &= \gamma_f \boldsymbol{\mu}^T \mathbf{W} \mathbf{I}(0) \\ T_n &= \gamma_n \boldsymbol{\mu}^T \mathbf{W} \mathbf{I}(\tau_0). \end{aligned} \qquad (38)$$

In summary, by assuming known eigenvalues and eigenvectors for the Jacobian of the first order system of equations resulting from the directional discretization of the RTE, an analytical solution has emerged. The solution exhibits several features not commonly found in previous solutions. In particular, an explicit form expressed as matrix functions results-- unlike most solutions to date, where one simply states that an algebraic system for the coefficients of the exponential eigensolutions is to be solved [20,21]. In addition, the solution is universally applicable even for the conservative case and for any source for which the integrals in the particular solution of Eq(25a) exist. Most importantly from a pedagogical perspective, an analytical and numerically stable solution comes about from classical mathematics, requiring no ad-hoc procedures for numerical evaluation as is now shown. Finally, it should be noted that similar forms for the exiting intensities have been found (e.g.,[16]), but not in the consistent manner presented here.

### III. Numerical Demonstration for a Homogeneous Medium

Numerical implementation of Eqs(32), (37) and (38) is through a FORTRAN program and will cover isotropic and beam incidence. In both cases, comparison to benchmarks found in the literature and internal conservation provides verification of the general formulation and claimed accuracy. Several additional features are to be included in the numerical evaluation–– the "faux quadrature" to find the intensity in any direction and position, the special case of $\mu = 0$ and convergence acceleration to best achieve extreme accuracy.

### III.1. Isotropic incidence

The most straightforward case to consider first is an isotropically entering intensity, where one assumes a uniform intensity distribution (normalized to unity) to enter the top surface and none at the bottom surface

$$I^+(0) = \mathbf{1}$$
$$I^-(\tau_0) = \mathbf{0}.$$

From Eqs(32) and (33) therefore

$$\begin{bmatrix} I^-(0) \\ I^+(\tau_0) \end{bmatrix} = R \begin{bmatrix} \mathbf{0} \\ \mathbf{1} \end{bmatrix} = \begin{bmatrix} R_f \mathbf{1} \\ T_n \mathbf{1} \end{bmatrix}, \qquad (39a)$$

and the angular intensity from Eq(37) becomes

$$\begin{bmatrix} I^-(\tau) \\ I^+(\tau) \end{bmatrix} = \frac{1}{2} \left\{ \begin{bmatrix} A^+(\tau-\tau_0) & A^+(\tau) \\ A^-(\tau-\tau_0) & A^-(\tau) \end{bmatrix} R + \begin{bmatrix} A^+(\tau) & A^+(\tau-\tau_0) \\ A^-(\tau) & A^-(\tau-\tau_0) \end{bmatrix} \right\} \begin{bmatrix} 0 \\ 1 \end{bmatrix}. \quad (39b)$$

At this point, we have established the approximate intensity $I(\tau;N)$ vector, which is the $N^{th}$ order angular approximation to the exact intensity. For example, the true reflectance is the limit as $N$ approaches infinity of the partial sums

$$R_f = \lim_{N \to \infty} R_f(N) = \gamma_f \lim_{N \to \infty} \sum_{m=1}^{2N} \omega_m(N) \mu_m(N) I_m(0;N), \quad (40)$$

where quadrature weight and abscissae dependence on $N$ is included to indicate the full extent of the limit. Thus, one observes the reflectance to be a sequence of reflectances whose limit is the true solution. The concept of convergence acceleration is to replace the original sequence by a new one, $\hat{R}_f(N)$, that converges more rapidly such that as $N \to \infty$

$$\frac{R_f - \hat{R}_f(N)}{R_f - R_f(N)} \to 0. \quad (41)$$

There are literally an infinity of choices for the alternative sequence. Here, we choose the Wynn-epsilon (*W-e*) algorithm [19], which has successfully accelerated a wide variety of sequences without requiring regularity in $N$. The *W-e* algorithm is

$$\begin{aligned} \varepsilon_{-1}^{(N)} &\equiv 0 \\ \varepsilon_0^{(N)} &\equiv R_f(N), \ N = 1, 2, \ldots \\ \varepsilon_{k+1}^{(N)} &= \varepsilon_{k-1}^{(N+1)} + \left[ \varepsilon_k^{(N+1)} - \varepsilon_k^{(N)} \right]^{-1}, \ k = 0, \ldots, 2K-1 \ ; \ N = 1, 2, \ldots, \end{aligned} \quad (42)$$

where $\varepsilon_0^{(N)}$ is the sequence to accelerate. The algorithm forms the following tableau:

$$\begin{array}{llllll} \varepsilon_0^{(1)} & \varepsilon_1^{(1)} & \varepsilon_2^{(1)} & \varepsilon_3^{(1)} & \varepsilon_4^{(1)} & \varepsilon_4^{(1)} \\ \varepsilon_0^{(2)} & \varepsilon_1^{(2)} & \varepsilon_2^{(2)} & \varepsilon_3^{(2)} & \varepsilon_4^{(2)} \\ \varepsilon_0^{(3)} & \varepsilon_1^{(3)} & \varepsilon_2^{(3)} & \varepsilon_3^{(3)} \\ \varepsilon_0^{(4)} & \varepsilon_1^{(4)} & \varepsilon_2^{(4)} \\ \varepsilon_0^{(5)} & \varepsilon_1^{(5)} \\ \varepsilon_0^{(6)} \end{array}$$

where the first column is the original sequence and every second column thereafter is potentially an accelerated approximation to the limit. Typically, the bottom diagonal will be the fastest to converge. To create the original sequence of reflectances and transmittances in $N$, the approximations generally proceed from an initial angular discretization $N_0$ to convergence by increments, of, say 4. One bases convergence on the relative error between every other element along the bottom diagonal of the tableau. Also, a window of only the last 10 elements of the original sequence participate in the acceleration at each $N$. Note that $N_0$ can be adjusted to reduce the overall computational time as desired by choosing a value near convergence.

A first verification considers a Mie scattering benchmark for the ($L = 8$) with the phase function given by the following scattering coefficients [6]:

Table 1
$L = 8$ Mie Scattering Phase Function

| $l$ | $\omega_l$ |
|---|---|
| 0 | 1.00000 |
| 1 | 2.00916 |
| 2 | 1.56339 |
| 3 | 0.67407 |
| 4 | 0.22215 |
| 5 | 0.04725 |
| 6 | 0.00671 |
| 7 | 0.00068 |
| 8 | 0.00005 |

Table 2a gives the reflectance and transmittance for slabs of two thickness each with $\omega$ varying over 0.9, 0.99 and 0.999. The results are in complete agreement with those of the benchmark [6]. To provide a more complete benchmark, we extend the results by two digits and include the total $R_f+T_n$ for later use. Also included in the table is the quadrature order $N$ at convergence. Negative entries indicate convergence by $W$-$e$; while, positive the original sequence converged first. One notes the advantage of acceleration by comparison of the last two columns of quadrature order with and without acceleration. The gain through acceleration, modest as it is, seems to be primarily for thin slabs. To add to this benchmark, Table 2b includes the reflectance and transmittance for a conservative medium of slab thicknesses varying from 0.01 to 1000 optical depths. For this case, only the internal consistency of the sum of reflectance and transmittance to unity is available, which we observe to 6- places. Again, the advantage is greatest for the thin slabs.

Figure 1 gives a graphical demonstration of the ability of the $W$-$e$ acceleration to achieve true extreme accuracy where, for a conservative slab of unit thickness, the reflectance and transmittance converged to 12- places. As the quadrature order increases, the figure shows the ratio of the relative errors with and without acceleration. The ability of *W-e* to out perform the original by several orders of magnitude at the extreme $10^{-12}$ relative error is apparent.

Table 2a
Comparison to Benchmark in [6]

| $\omega$ | $\tau_0$ | $R_f$ | $T_n$ | $R_f+T_n$ | $N$ | $N_{ori}$ |
|---|---|---|---|---|---|---|
| 9.000E-01 | 1.000E+00 | 1.719133E-01 | 6.542669E-01 | 8.261802E-01 | -22 | 26 |
| 9.000E-01 | 1.000E+01 | 2.907016E-01 | 3.293595E-02 | 3.236376E-01 | 22 | 22 |
| 9.900E-01 | 1.000E+00 | 2.266183E-01 | 7.536775E-01 | 9.802958E-01 | -22 | 26 |
| 9.900E-01 | 1.000E+01 | 6.220622E-01 | 2.107840E-01 | 8.328462E-01 | 22 | 22 |
| 9.990E-01 | 1.000E+00 | 2.331042E-01 | 7.648988E-01 | 9.980030E-01 | -22 | 26 |
| 9.990E-01 | 1.000E+01 | 7.069447E-01 | 2.734408E-01 | 9.803855E-01 | 18 | 18 |
| 9.999E-01 | 1.000E+00 | 2.337645E-01 | 7.660355E-01 | 9.998000E-01 | -22 | 26 |
| 9.999E-01 | 1.000E+01 | 7.169136E-01 | 2.810904E-01 | 9.980039E-01 | 18 | 18 |

Table 2b
Addition to Benchmark in [6]

| 1.000E+00 | 1.000E-02 | 4.672649E-03 | 9.953274E-01 | 1.000000E+00 | -78 | 86 |
| 1.000E+00 | 1.000E-01 | 3.945935E-02 | 9.605406E-01 | 1.000000E+00 | -42 | 46 |
| 1.000E+00 | 1.000E+00 | 2.338381E-01 | 7.661619E-01 | 1.000000E+00 | -22 | 26 |
| 1.000E+00 | 1.000E+01 | 7.180410E-01 | 2.819590E-01 | 1.000000E+00 | 18 | 18 |
| 1.000E+00 | 1.000E+02 | 9.613011E-01 | 3.869892E-02 | 1.000000E+00 | 14 | 14 |
| 1.000E+00 | 1.000E+03 | 9.959804E-01 | 4.019624E-03 | 1.000000E+00 | 14 | 14 |

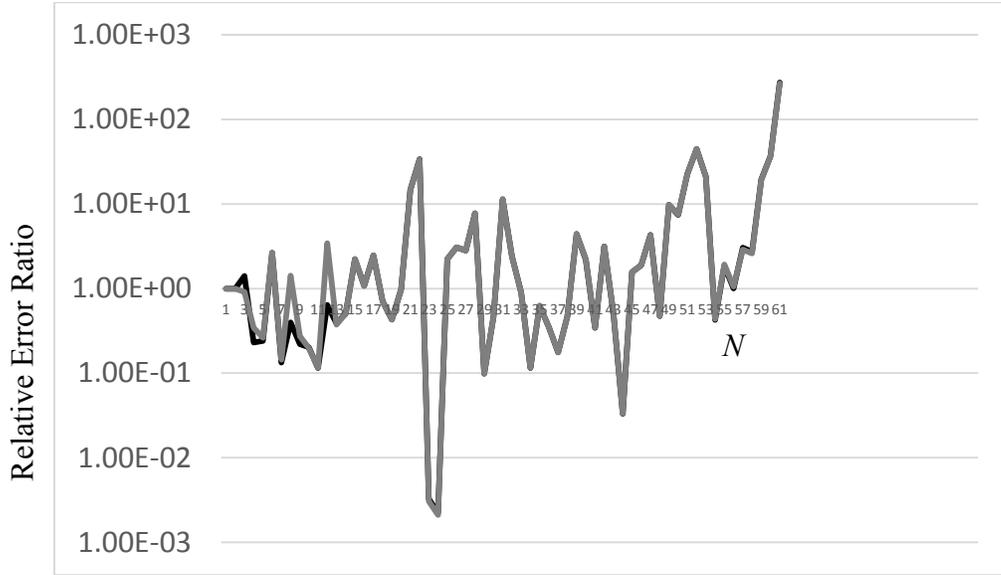

Fig. 1. Ratio of relative error between quadrature order approximations [without acceleration/with acceleration].
▬▬▬ Reflectance
▬▬▬ Transmittance

As further verification, Table 3 gives the interior flux approximation in a conservative medium of unit thickness for requested relative errors of $10^{-4(1)8}$ between consecutive $N^{th}$ order approximations. The flux is spatially uniform for this case according to the following analysis. If the flux is

$$J(\tau) \equiv \int_{-1}^{1} d\mu \mu I(\tau,\mu), \qquad (43)$$

then upon integration of Eq(1) over all directions without a fixed source

$$\frac{dJ(\tau)}{d\tau} = 0,$$

which in turn implies

$$J(\tau) = J(0) = \frac{1}{2}(1 - R_f) = \frac{1}{2}T_n \qquad (44a)$$

with

$$R_f \equiv \int_{-1}^{0} d\mu \mu I(0,\mu) / \int_{0}^{1} d\mu \mu I(0,\mu)$$

$$T_n \equiv \int_{0}^{1} d\mu \mu I(\tau_0,\mu) / \int_{0}^{1} d\mu \mu I(0,\mu). \qquad (44b)$$

Even for a relative error of $10^{-4}$, the flux is spatially uniform to 5- places and for error of $10^{-8}$ to 8-places. $N$ at convergence is in parenthesis with both the original accelerated sequence essentially converging simultaneously. This comparison gives further confidence in the accuracy of the RM/DOM algorithm.

Before considering beam incidence, we end this section on a computational note. Above, it was mention that Eq(32b) was preferable over Eq(32a). The reason is one of numerical stability when the eigenvalue $\lambda_k$ becomes large. To use Eq(32a), the inverse of the matrix on the LHS is necessary and therefore will require $\boldsymbol{B}^{-1}$. However, it is apparent that for large $\lambda_k$

$$B \sim T\left[diag\left\{\lambda_k e^{-\lambda_k \tau_0}\right\}\right]T^{-1}.$$

Table 3
Flux for spatial variation of flux ($\tau_0 = 1$, $\omega = 1$)

| $\tau$ | $10^{-4}$ (9) | $10^{-5}$ (13) | $10^{-6}$ (13) | $10^{-7}$ (17) | $10^{-8}$(21) |
|---|---|---|---|---|---|
| 0.00E+00 | 3.83080683E-01 | 3.83080959E-01 | 3.83080959E-01 | 3.83080969E-01 | 3.83080971E-01 |
| 1.00E-01 | 3.83080831E-01 | 3.83080959E-01 | 3.83080959E-01 | 3.83080969E-01 | 3.83080971E-01 |
| 2.00E-01 | 3.83080751E-01 | 3.83080959E-01 | 3.83080959E-01 | 3.83080969E-01 | 3.83080971E-01 |
| 3.00E-01 | 3.83080740E-01 | 3.83080959E-01 | 3.83080959E-01 | 3.83080969E-01 | 3.83080971E-01 |
| 4.00E-01 | 3.83080736E-01 | 3.83080959E-01 | 3.83080959E-01 | 3.83080969E-01 | 3.83080971E-01 |
| 5.00E-01 | 3.83080734E-01 | 3.83080959E-01 | 3.83080959E-01 | 3.83080969E-01 | 3.83080971E-01 |
| 6.00E-01 | 3.83080733E-01 | 3.83080959E-01 | 3.83080959E-01 | 3.83080969E-01 | 3.83080971E-01 |
| 7.00E-01 | 3.83080732E-01 | 3.83080959E-01 | 3.83080959E-01 | 3.83080969E-01 | 3.83080971E-01 |
| 8.00E-01 | 3.83080731E-01 | 3.83080959E-01 | 3.83080959E-01 | 3.83080969E-01 | 3.83080971E-01 |
| 9.00E-01 | 3.83080731E-01 | 3.83080959E-01 | 3.83080959E-01 | 3.83080969E-01 | 3.83080971E-01 |
| 1.00E+00 | 3.83080730E-01 | 3.83080959E-01 | 3.83080959E-01 | 3.83080969E-01 | 3.83080971E-01 |

Hence, when the exponential experiences numerical underflow, the matrix $B$ becomes numerically singular $|B|=0$ and one cannot solve Eq(32a). For this reason, rearrangement into Eq(32b), which is solvable, is the only useful from of Eq(32a). In addition, since we know the intensities on the RHS, Eq (32b) makes perfect physical sense; while, Eq(32a) does not.

### III.2. Beam incidence

For a beam in direction $\mu_0$ incident on the top surface, the boundary conditions are

$$I(0,\mu) = I_{inc}\delta(\mu-\mu_0), \ \mu > 0$$
$$I(\tau_0,\mu) = 0, \mu < 0. \tag{45}$$

As is common practice, one seeks to reformulate the beam source as a volume source and then to solve Eq(4a). By decomposing the flux into uncollided and collided contributions

$$I(\tau,\mu) = I_0(\tau,\mu) + I_c(\tau,\mu), \tag{46a}$$

the uncollided photons obey the transport equation with the surface source but no collisions

$$\left[\mu\frac{\partial}{\partial \tau} + 1\right]I_0(\tau,\mu) = 0$$
$$I_0(0,\mu) = I_{inc}\delta(\mu-\mu_0), \ \mu,\mu_0 > 0 \tag{46b}$$
$$I_0(\tau_0,\mu) = 0, \mu < 0.$$

The collided photons obey the transport equation with the uncollided contribution explicitly included in the scattering term (now a volume source) but with homogeneous boundary conditions

$$\left[\mu\frac{\partial}{\partial\tau}+1\right]I_c(\tau,\mu)=\omega\int_{-1}^{1}d\mu'f(\mu',\mu)I_c(\tau,\mu')+\omega\int_{-1}^{1}d\mu'f(\mu',\mu)I_0(\tau,\mu')$$

$$I_c(0,\mu)=0,\ \mu>0 \tag{46c}$$

$$I_c(\tau_0,\mu)=0,\ \mu<0.$$

Since the solution for the uncollided intensity is

$$I_0(\tau,\mu)=I_{inc}e^{-\tau/\mu_0}\delta(\mu-\mu_0), \tag{47}$$

Eq(46c) with a known volume source becomes

$$\left[\mu\frac{\partial}{\partial\tau}+1\right]I(\tau,\mu)=\omega\int_{-1}^{1}d\mu'f(\mu',\mu)I(\tau,\mu')+\omega I_{inc}e^{-\tau/\mu_0}f(\mu_0,\mu). \tag{48}$$

Note that we have suppressed the subscript "$c$" and the dependence on $\mu_0$. One can now determine the following quantities required to obtain the angular intensity as specified above by Eqs(6d) and (10b,c)

$$Q^{\pm}(\tau)\equiv\frac{\omega}{2}I_{inc}\sum_{l=0}^{L}\beta_l P_l(\mu_0)\boldsymbol{P}_l^{\pm}e^{-\tau/\mu_0}$$

$$\boldsymbol{\xi}^{\pm}(\tau)=\boldsymbol{\xi}_0^{\pm}e^{-\tau/\mu_0} \tag{49a,b,c}$$

$$\boldsymbol{q}(\tau)\equiv-\boldsymbol{u}_0 e^{-\tau/\mu_0},$$

where

$$\boldsymbol{P}_l^{\pm}\equiv\left[P_{l\genfrac{\{}{\}}{0pt}{}{1}{N+1}}\ \ P_{l\genfrac{\{}{\}}{0pt}{}{2}{N+2}}\ \ \ldots\ \ P_{l\genfrac{\{}{\}}{0pt}{}{N}{2N}}\right]^T$$

$$\boldsymbol{\xi}_0^{\pm}=\frac{\omega}{2}I_{inc}\sum_{l=0}^{L}\beta_l P_l(\mu_0)\left[1\pm(-1)^l\right]\boldsymbol{M}^{-1}\boldsymbol{P}_l \tag{49d,e,f}$$

$$u_0 \equiv \frac{\xi_0^-}{\mu_0} + (\alpha + \beta)\xi_0^+.$$

For the particular solution therefore, we require the solution to the inhomogeneous ODE

$$\left[\frac{d^2}{d\tau^2} - \lambda_k^2\right]\Theta_{Pk}(\tau) = -(Tu_0)_k e^{-\tau/\mu_0}. \tag{50a}$$

Most simply, if

$$\Theta_{Pk}(\tau) = a_k e^{-\tau/\mu_0},$$

then for $\lambda_k \neq 1/\mu_0$

$$a_k = -\frac{\mu_0^2}{1 - \mu_0^2 \lambda_k^2}(T^{-1}u_0)_k \tag{50b}$$

and therefore

$$\Theta_P(\tau) = -diag\left\{\frac{\mu_0^2}{1 - \mu_0^2 \lambda_0^2}\right\}T^{-1}u_0 e^{-\tau/\mu_0} \tag{50c}$$

giving the particular solution

$$\psi_P^+(\tau) = -Tdiag\left\{\frac{\mu_0^2}{1 - \mu_0^2 \lambda_k^2}\right\}T^{-1}u_0 e^{-\tau/\mu_0}. \tag{50d}$$

For the singular case $\lambda_k = 1/\mu_0$, we assume a solution

$$\Theta_{Pk}(\tau) = a_k \tau e^{-\tau/\mu_0}$$

to give

$$a_k = \frac{\mu_0}{2}\left(\bm{T}^{-1}\bm{u}_0\right)_k. \tag{50e}$$

Hence, the final form of the particular solution is

$$\bm{\psi}_P^+(\tau,\mu_0) = \bm{R}_P(\tau,\mu_0) \equiv -\bm{T}\,diag\left\{\begin{array}{l}-\dfrac{\mu_0}{2}\tau;\ \lambda_k = 1/\mu_0 \\[1ex] \dfrac{\mu_0^2}{1-\mu_0^2\lambda_k^2};\ \lambda_k \neq 1/\mu_0\end{array}\right\}\bm{T}^{-1}\bm{u}_0 e^{-\tau/\mu_0}. \tag{51a}$$

Note that the particular solution as stated does not vanish at the slab surfaces. By collecting all particular solutions in Eq(26a) into a single term, one can define the particular solution to appropriately vanish as

$$\bm{\psi}_P^+(\tau) \to \bm{R}_P(\tau,\mu_0) - H(\tau)\bm{R}_P(\tau_0,\mu_0) - H(\tau_0-\tau)\bm{R}_P(0,\mu_0). \tag{51b}$$

As an exercise in integration, we find the identical result in the Appendix from the more general form of the particular solution given by Eq(25a).

We are now in position to numerically evaluate the angular flux, but before doing so, let us consider the role of convergence acceleration in determining the angular intensity and how we are to address determination of the intensity at $\mu = 0$.

To accelerate the angular intensity in a particular (edit) direction requires that the intensity in that direction be determined at each quadrature order $N$. In this way, the edit will form a (original) sequence in $N$, as $N$ is incremented. This is commonly done by inserting the known scattering term at each $N$ (a sum over all angular intensities) into the radiative transfer equation [Eq(4a)] for the desired edit direction and solve as a first order ODE {e.g., see [21]}. However, one can achieve the identical result much more easily.

In particular, it is more straightforward to simply add the desired angular edits to the quadrature list with zero weight, called a "faux quadrature". Now the transport equation itself at each $N$ interpolates the angular intensity at the edit points without disturbing the angular intensities at the $N$ quadrature points. Note that this procedure requires the singular case for the particular solution if one of the edits happens to be $\mu_0$.

A secondary consideration is how one addresses the case $\mu = 0$. For $0 < \tau < \tau_0$ and $\mu_z = 0$ in Eq(4a), we find the angular intensity satisfies

$$I_z(\tau) = \omega \sum_{m'=1}^{2N} \omega_{m'} f(\mu_{m'}, 0) I_{m'}(\tau) + Q(\tau, 0). \tag{52}$$

Thus, $I_z(\tau)$ comes from the angular intensity vector at the quadrature points. At the top and bottom surfaces, letting $\mu$ approach zero from the negative and positive directions respectively gives the same result.

Incorporating these two numerical procedures into the evaluation of Eq(37) will then give the angular intensity in any direction and optical depth. Note that for the results to follow, we expect all digits to be correct to within one unit in the last place reported.

To begin the computation, we require the exiting angular intensities from Eq(32b), which are

$$\begin{bmatrix} I^-(0) \\ I^+(\tau_0) \end{bmatrix} \equiv \begin{bmatrix} \left\{ (x^- - B)^{-1} + (x^- + B)^{-1} \right\} & \left\{ (x^- - B)^{-1} - (x^- + B)^{-1} \right\} \\ \left\{ (x^- - B)^{-1} - (x^- + B)^{-1} \right\} & \left\{ (x^- - B)^{-1} + (x^- + B)^{-1} \right\} \end{bmatrix} \begin{bmatrix} \left. \frac{d\psi_P^+(\tau)}{d\tau} \right|_{\tau=0} + \xi^-(0) \\ \left. \frac{d\psi_P^+(\tau)}{d\tau} \right|_{\tau=\tau_0} + \xi^-(\tau_0) \end{bmatrix}. \tag{53}$$

Initial verification of the angular intensity comes from the Mie scattering benchmark of the last section [7]. The spatial edits are on the optical depths $\tau = \tau_0 / s$, $s = 20, 10, 5, 2, 4/3, 1$ and angular edits at $\mu = -1(0.1)1$. Table 4a shows the angular intensity for $\omega = 0.95$, $\mu_0 = 0.5$ and $\tau_0 = 1$. On rounding, all digits agree to all five places of the benchmark. Two additional digits are included to provide additional usefulness.

Table 4a
$L = 8$ Mie Scatter Kernel for $m = 0$
($\omega = 0.95$, $\mu_0 = 0.5$, $\tau_0 = 1$, $I_{inc} = 0.5$)

| $\mu \backslash \tau$ | 0 | $\tau_0/20$ | $\tau_0/10$ | $\tau_0/5$ | $\tau_0/2$ | $3\tau_0/4$ | $\tau_0$ |
|---|---|---|---|---|---|---|---|
| -1.000E+00 | 4.7680739E-02 | 4.4191232E-02 | 4.0646720E-02 | 3.3709854E-02 | 1.5857241E-02 | 5.4529708E-03 | 0.0000000E+00 |
| -9.000E-01 | 6.4564440E-02 | 6.0374289E-02 | 5.6013853E-02 | 4.7289947E-02 | 2.3812095E-02 | 9.0108349E-03 | 0.0000000E+00 |
| -8.000E-01 | 8.4587655E-02 | 7.9676269E-02 | 7.4442518E-02 | 6.3753956E-02 | 3.3828695E-02 | 1.3691858E-02 | 0.0000000E+00 |
| -7.000E-01 | 1.0834976E-01 | 1.0271869E-01 | 9.6567817E-02 | 8.3748795E-02 | 4.6491789E-02 | 1.9884482E-02 | 0.0000000E+00 |
| -6.000E-01 | 1.3650449E-01 | 1.3020361E-01 | 1.2312760E-01 | 1.0806281E-01 | 6.2604846E-02 | 2.8172141E-02 | 0.0000000E+00 |
| -5.000E-01 | 1.6967721E-01 | 1.6285165E-01 | 1.5491892E-01 | 1.3761804E-01 | 8.3292056E-02 | 3.9479159E-02 | 0.0000000E+00 |
| -4.000E-01 | 2.0822991E-01 | 2.0120060E-01 | 1.9262576E-01 | 1.7335796E-01 | 1.1012425E-01 | 5.5363056E-02 | 0.0000000E+00 |
| -3.000E-01 | 2.5167744E-01 | 2.4506518E-01 | 2.3630972E-01 | 2.1580644E-01 | 1.4514429E-01 | 7.8629773E-02 | 0.0000000E+00 |
| -2.000E-01 | 2.9752321E-01 | 2.9240107E-01 | 2.8424081E-01 | 2.6379979E-01 | 1.9003527E-01 | 1.1455737E-01 | 0.0000000E+00 |
| -1.000E-01 | 3.4012557E-01 | 3.3847566E-01 | 3.3198375E-01 | 3.1288756E-01 | 2.4121715E-01 | 1.7069703E-01 | 0.0000000E+00 |
| 0.000E+00 | 3.5937904E-01 | 3.7448490E-01 | 3.7380021E-01 | 3.5934745E-01 | 2.8825763E-01 | 2.2562269E-01 | 0.0000000E+00 |
| 0.000E+00 | 0.0000000E+00 | 3.7448490E-01 | 3.7380021E-01 | 3.5934745E-01 | 2.8825763E-01 | 2.2562269E-01 | 1.5152044E-01 |
| 1.000E-01 | 0.0000000E+00 | 1.5561969E-01 | 2.5142670E-01 | 3.3886425E-01 | 3.3070258E-01 | 2.6985545E-01 | 2.0307340E-01 |
| 2.000E-01 | 0.0000000E+00 | 9.2437025E-02 | 1.6508021E-01 | 2.6188673E-01 | 3.3551731E-01 | 3.0155334E-01 | 2.4424153E-01 |
| 3.000E-01 | 0.0000000E+00 | 6.7072588E-02 | 1.2414425E-01 | 2.1038294E-01 | 3.1480528E-01 | 3.0971262E-01 | 2.7030221E-01 |
| 4.000E-01 | 0.0000000E+00 | 5.3016958E-02 | 9.9893927E-02 | 1.7530448E-01 | 2.8818268E-01 | 3.0293863E-01 | 2.8106600E-01 |
| 5.000E-01 | 0.0000000E+00 | 4.3718197E-02 | 8.3238337E-02 | 1.4925136E-01 | 2.6130432E-01 | 2.8846360E-01 | 2.8082243E-01 |
| 6.000E-01 | 0.0000000E+00 | 3.6786430E-02 | 7.0519358E-02 | 1.2832314E-01 | 2.3525903E-01 | 2.6979273E-01 | 2.7302917E-01 |
| 7.000E-01 | 0.0000000E+00 | 3.1141106E-02 | 5.9990823E-02 | 1.1038662E-01 | 2.0993897E-01 | 2.4847557E-01 | 2.5982225E-01 |
| 8.000E-01 | 0.0000000E+00 | 2.6222496E-02 | 5.0716634E-02 | 9.4200130E-02 | 1.8498212E-01 | 2.2517903E-01 | 2.4245347E-01 |
| 9.000E-01 | 0.0000000E+00 | 2.1713290E-02 | 4.2153804E-02 | 7.9003938E-02 | 1.6003444E-01 | 2.0018674E-01 | 2.2166696E-01 |
| 1.000E+00 | 0.0000000E+00 | 1.7423141E-02 | 3.3971636E-02 | 6.4319263E-02 | 1.3481854E-01 | 1.7362739E-01 | 1.9793246E-01 |

Table 4b
$L = 8$ Mie Scatter Kernel for $m = 8$
($\omega = 0.95$, $\mu_0 = 0.5$, $\tau_0 = 1$, $I_{inc} = 1$)

| $\mu \backslash \tau$ | 0 | $\tau_0/20$ | $\tau_0/10$ | $\tau_0/5$ | $\tau_0/2$ | $3\tau_0/4$ | $\tau_0$ |
|---|---|---|---|---|---|---|---|
| -1.000E+00 | 0.0000000E+00 | 0.0000000E+00 | 0.0000000E+00 | 0.0000000E+00 | 0.0000000E+00 | 0.0000000E+00 | 0.0000000E+00 |
| -9.000E-01 | 6.5625391E-10 | 5.8914304E-10 | 5.2815237E-10 | 4.2219714E-10 | 1.9934608E-10 | 8.2847308E-11 | 0.0000000E+00 |
| -8.000E-01 | 9.1636878E-09 | 8.2326319E-09 | 7.3863711E-09 | 5.9157513E-09 | 2.8165273E-09 | 1.1832473E-09 | 0.0000000E+00 |
| -7.000E-01 | 4.0249326E-08 | 3.6190815E-08 | 3.2501619E-08 | 2.6088956E-08 | 1.2547246E-08 | 5.3429052E-09 | 0.0000000E+00 |
| -6.000E-01 | 1.0966293E-07 | 9.8703386E-08 | 8.8741262E-08 | 7.1422850E-08 | 3.4781849E-08 | 1.5070373E-08 | 0.0000000E+00 |
| -5.000E-01 | 2.2918971E-07 | 2.0652282E-07 | 1.8592286E-07 | 1.5011771E-07 | 7.4263736E-08 | 3.2929240E-08 | 0.0000000E+00 |
| -4.000E-01 | 4.0364654E-07 | 3.6419928E-07 | 3.2836806E-07 | 2.6613595E-07 | 1.3433461E-07 | 6.1509004E-08 | 0.0000000E+00 |
| -3.000E-01 | 6.2943646E-07 | 5.6869337E-07 | 5.1357757E-07 | 4.1802334E-07 | 2.1651285E-07 | 1.0392686E-07 | 0.0000000E+00 |
| -2.000E-01 | 8.9447618E-07 | 8.0904612E-07 | 7.3165782E-07 | 5.9791370E-07 | 3.1941422E-07 | 1.6505266E-07 | 0.0000000E+00 |
| -1.000E-01 | 1.1813127E-06 | 1.0688908E-06 | 9.6716367E-07 | 7.9180936E-07 | 4.3350657E-07 | 2.5046511E-07 | 0.0000000E+00 |
| 0.000E+00 | 1.4757328E-06 | 1.3352990E-06 | 1.2082288E-06 | 9.8921442E-07 | 5.4289266E-07 | 3.2928109E-07 | 0.0000000E+00 |
| 0.000E+00 | 0.0000000E+00 | 1.3352990E-06 | 1.2082288E-06 | 9.8921442E-07 | 5.4289266E-07 | 3.2928109E-07 | 1.9971892E-07 |
| 1.000E-01 | 0.0000000E+00 | 5.2859344E-07 | 7.9889942E-07 | 9.4798249E-07 | 6.3993602E-07 | 3.9440254E-07 | 2.3973118E-07 |
| 2.000E-01 | 0.0000000E+00 | 2.6329242E-07 | 4.4328927E-07 | 6.3180326E-07 | 5.9702947E-07 | 4.1699385E-07 | 2.6864210E-07 |
| 3.000E-01 | 0.0000000E+00 | 1.4763750E-07 | 2.5856043E-07 | 3.9695812E-07 | 4.5287275E-07 | 3.5683886E-07 | 2.5213926E-07 |
| 4.000E-01 | 0.0000000E+00 | 8.2070715E-08 | 1.4668783E-07 | 2.3433850E-07 | 2.9894033E-07 | 2.5632829E-07 | 1.9562189E-07 |
| 5.000E-01 | 0.0000000E+00 | 4.2249686E-08 | 7.6458207E-08 | 1.2519740E-07 | 1.7177455E-07 | 1.5627982E-07 | 1.2638468E-07 |
| 6.000E-01 | 0.0000000E+00 | 1.8825271E-08 | 3.4353900E-08 | 5.7206557E-08 | 8.2593552E-08 | 7.8453748E-08 | 6.6279496E-08 |
| 7.000E-01 | 0.0000000E+00 | 6.5456093E-09 | 1.2017087E-08 | 2.0256111E-08 | 3.0365796E-08 | 2.9798350E-08 | 2.6036280E-08 |
| 8.000E-01 | 0.0000000E+00 | 1.4283575E-09 | 2.6342495E-09 | 4.4814591E-09 | 6.9147374E-09 | 6.9599115E-09 | 6.2449834E-09 |
| 9.000E-01 | 0.0000000E+00 | 9.8856917E-11 | 1.8296411E-10 | 3.1352189E-10 | 4.9491896E-10 | 5.0836713E-10 | 4.6603207E-10 |
| 1.000E+00 | 0.0000000E+00 | 0.0000000E+00 | 0.0000000E+00 | 0.0000000E+00 | 0.0000000E+00 | 0.0000000E+00 | 0.0000000E+00 |

While our focus has been exclusively on the azimuthally integrated angular intensity, we now show that the analysis also yields accurate results for Fourier moments {see [7] or [20] for further details}. To include an $m^{th}$ Fourier component, the following two modifications in the above derivation are necessary:

$$P_l(\mu) \to P_l^m(\mu)$$ (Legendre polynomials go to the associated Legendre functions)

and

$$\beta_l \to \frac{(l-m)!}{(l-m)!}\beta_l.$$

With these changes and the beam normalized to unity, Table 4b presents results for $m = 8$. As for the azimuthally integrated angular intensity, 5 of the 7 digits quoted agree to all 5- digits of the previously referenced benchmark. The same agreement is true for components $m = 1$ through 7 (not shown). A more comprehensive treatment for the general case of Fourier components utilizing the advantage of parallel computing (since all components are independent) will be the subject of a future effort.

The last two cases we consider originate from those designated by the Radiation Commission of the International Association of Meteorology and Atmospheric Physics [13] as appropriate test problems for radiative transfer code development and are severe tests of any numerical method. The first is for the HAZE-L kernel, a modestly forward peaked scattering phase function characteristic of clouds. For $L = 82$, the HAZE-L scattering coefficients are given in Table 5a with the angular intensities to 7-places for $\omega = 1$ and 0.9 in a homogenous medium of thickness unity given in Tables 5b,c. On rounding, the first five places are in complete agreement with those of the benchmark [7].

The scattering coefficients for second case of the Cloud $C_1$ phase function, also characteristic of clouds but more forward peaked, for $\tau_0 = 64$, are given in Tables 6a with the angular intensities to 7- digits given in Tables 6b,c for $\omega = 1$ and 0.9. One observes the same agreement with the published benchmark [7] as in the previous cases.

Table 7 summarizes the quadrature orders required for the RM/DOM to produce the 7-digit benchmarks of Tables 5b,c and 6b,c. Unexpectedly, conservative HAZE-L scattering required the largest quadrature order most probably because the medium is relatively thin. The quadrature order required for 5-place accuracy for the Cloud $C_1$ case is generally less than the ADO benchmark of Siewert [20], which is ~350.

Table 5a
Scattering Coefficients for HAZE-L phase function

| l | $\beta_1$ | $\beta_{1+16}$ | $\beta_{1+32}$ | $\beta_{1+48}$ | $\beta_{1+64}$ | $\beta_{1+80}$ |
|---|---|---|---|---|---|---|
| 0 | 1 | 0.34688 | 0.01711 | 0.00107 | 0.00008 | 0.00001 |
| 1 | 2.41260 | 0.28351 | 0.01298 | 0.00082 | 0.00006 | 0.00001 |
| 2 | 3.23047 | 0.23317 | 0.01198 | 0.00077 | 0.00006 | 0.00001 |
| 3 | 3.37296 | 0.18963 | 0.0904 | 0.00059 | 0.00005 | |
| 4 | 3.23150 | 0.15788 | 0.00841 | 0.00055 | 0.00004 | |
| 5 | 2.89350 | 0.12739 | 0.00634 | 0.00043 | 0.00004 | |
| 6 | 2.49594 | 0.10762 | 0.00592 | 0.00040 | 0.00003 | |
| 7 | 2.11361 | 0.08597 | 0.00446 | 0.00031 | 0.00003 | |
| 8 | 1.74812 | 0.07381 | 0.00416 | 0.00029 | 0.00002 | |
| 9 | 1.44692 | 0.05828 | 0.00316 | 0.00023 | 0.00002 | |
| 10 | 1.17714 | 0.05089 | 0.00296 | 0.00021 | 0.00002 | |
| 11 | 0.96643 | 0.03971 | 0.00225 | 0.00017 | 0.00001 | |
| 12 | 0.78237 | 0.03524 | 0.00210 | 0.00015 | 0.00001 | |
| 13 | 0.64114 | 0.02720 | 0.00160 | 0.00012 | 0.00001 | |
| 14 | 0.51966 | 0.02451 | 0.00150 | 0.00011 | 0.00001 | |
| 15 | 0.42563 | 0.01874 | 0.00115 | 0.00009 | 0.00001 | |

Table 5b

Angular Intensity for HAZE-L Scattering Kernel ($L = 82$)

($\omega = 1$, $\mu_0 = 1$, $\tau_0 = 1$, $I_{inc} = 0.5$)

| $\mu \backslash \tau$ | 0 | $\tau_0/20$ | $\tau_0/10$ | $\tau_0/5$ | $\tau_0/2$ | $3\tau_0/4$ | $\tau_0$ |
|---|---|---|---|---|---|---|---|
| -1.000E+00 | 3.6145156E-02 | 3.4339396E-02 | 3.2510866E-02 | 2.8812216E-02 | 1.7628611E-02 | 8.5258908E-03 | 0.0000000E+00 |
| -9.000E-01 | 3.9781870E-02 | 3.7872320E-02 | 3.5920682E-02 | 3.1930313E-02 | 1.9620173E-02 | 9.4573134E-03 | 0.0000000E+00 |
| -8.000E-01 | 4.2731263E-02 | 4.0840607E-02 | 3.8873442E-02 | 3.4767734E-02 | 2.1601856E-02 | 1.0395857E-02 | 0.0000000E+00 |
| -7.000E-01 | 4.8005147E-02 | 4.6131929E-02 | 4.4130697E-02 | 3.9829198E-02 | 2.5247889E-02 | 1.2217079E-02 | 0.0000000E+00 |
| -6.000E-01 | 5.5821353E-02 | 5.4043177E-02 | 5.2059351E-02 | 4.7598583E-02 | 3.1183660E-02 | 1.5361836E-02 | 0.0000000E+00 |
| -5.000E-01 | 6.6094221E-02 | 6.4629636E-02 | 6.2844874E-02 | 5.8497071E-02 | 4.0273976E-02 | 2.0562127E-02 | 0.0000000E+00 |
| -4.000E-01 | 7.8148081E-02 | 7.7440255E-02 | 7.6250769E-02 | 7.2704873E-02 | 5.3729974E-02 | 2.9128534E-02 | 0.0000000E+00 |
| -3.000E-01 | 8.9968154E-02 | 9.0770642E-02 | 9.0878384E-02 | 8.9471128E-02 | 7.2964349E-02 | 4.3468799E-02 | 0.0000000E+00 |
| -2.000E-01 | 9.7081540E-02 | 1.0042085E-01 | 1.0278927E-01 | 1.0550594E-01 | 9.8377715E-02 | 6.7994924E-02 | 0.0000000E+00 |
| -1.000E-01 | 9.2932814E-02 | 9.9818714E-02 | 1.0519502E-01 | 1.1349749E-01 | 1.2403692E-01 | 1.0839912E-01 | 0.0000000E+00 |
| 0.000E+00 | 6.9877391E-02 | 8.4667310E-02 | 9.4166299E-02 | 1.0872694E-01 | 1.3576248E-01 | 1.4277947E-01 | 0.0000000E+00 |
| 0.000E+00 | 0.0000000E+00 | 8.4667310E-02 | 9.4166299E-02 | 1.0872694E-01 | 1.3576248E-01 | 1.4277947E-01 | 1.1480771E-01 |
| 1.000E-01 | 0.0000000E+00 | 2.9541820E-02 | 5.2434564E-02 | 8.4564915E-02 | 1.3509602E-01 | 1.5610649E-01 | 1.5697621E-01 |
| 2.000E-01 | 0.0000000E+00 | 1.6490681E-02 | 3.2281653E-02 | 6.0752687E-02 | 1.2435036E-01 | 1.5892546E-01 | 1.7681766E-01 |
| 3.000E-01 | 0.0000000E+00 | 1.2342100E-02 | 2.4848764E-02 | 4.9396789E-02 | 1.1481121E-01 | 1.5793652E-01 | 1.8830112E-01 |
| 4.000E-01 | 0.0000000E+00 | 1.1187938E-02 | 2.2644991E-02 | 4.5754673E-02 | 1.1226864E-01 | 1.6086203E-01 | 2.0001870E-01 |
| 5.000E-01 | 0.0000000E+00 | 1.1795943E-02 | 2.3790964E-02 | 4.8000306E-02 | 1.1907946E-01 | 1.7319074E-01 | 2.1963289E-01 |
| 6.000E-01 | 0.0000000E+00 | 1.4204907E-02 | 2.8458379E-02 | 5.6873102E-02 | 1.3905102E-01 | 2.0144487E-01 | 2.5598334E-01 |
| 7.000E-01 | 0.0000000E+00 | 1.9583294E-02 | 3.8924848E-02 | 7.6745368E-02 | 1.8200357E-01 | 2.5898644E-01 | 3.2512495E-01 |
| 8.000E-01 | 0.0000000E+00 | 3.1953231E-02 | 6.2942983E-02 | 1.2204484E-01 | 2.7718191E-01 | 3.8276705E-01 | 4.6865779E-01 |
| 9.000E-01 | 0.0000000E+00 | 6.8726703E-02 | 1.3391677E-01 | 2.5425935E-01 | 5.4460066E-01 | 7.1944669E-01 | 8.4608373E-01 |
| 1.000E+00 | 0.0000000E+00 | 3.6493954E-01 | 7.0026634E-01 | 1.2895497E+00 | 2.5225517E+00 | 3.0931861E+00 | 3.3809098E+00 |

Table 5c

Angular Intensity for HAZE-L Scattering Kernel ($L = 82$)

($\omega = 0.9$, $\mu_0 = 1$, $\tau_0 = 1$, $I_{inc} = 0.5$)

| $\mu \backslash \tau$ | 0 | $\tau_0/20$ | $\tau_0/10$ | $\tau_0/5$ | $\tau_0/2$ | $3\tau_0/4$ | $\tau_0$ |
|---|---|---|---|---|---|---|---|
| -1.000E+00 | 2.7971665E-02 | 2.6583431E-02 | 2.5179489E-02 | 2.2342144E-02 | 1.3751918E-02 | 6.7043863E-03 | 0.0000000E+00 |
| -9.000E-01 | 3.0180197E-02 | 2.8742728E-02 | 2.7276328E-02 | 2.4282402E-02 | 1.5036989E-02 | 7.3279307E-03 | 0.0000000E+00 |
| -8.000E-01 | 3.1447755E-02 | 3.0070750E-02 | 2.8641094E-02 | 2.5662054E-02 | 1.6096218E-02 | 7.8550379E-03 | 0.0000000E+00 |
| -7.000E-01 | 3.4383906E-02 | 3.3055747E-02 | 3.1640694E-02 | 2.8605980E-02 | 1.8314210E-02 | 9.0026264E-03 | 0.0000000E+00 |
| -6.000E-01 | 3.9130810E-02 | 3.7890987E-02 | 3.6513506E-02 | 3.3427829E-02 | 2.2097749E-02 | 1.1061916E-02 | 0.0000000E+00 |
| -5.000E-01 | 4.5637920E-02 | 4.4617111E-02 | 4.3383966E-02 | 4.0403191E-02 | 2.8008565E-02 | 1.4514343E-02 | 0.0000000E+00 |
| -4.000E-01 | 5.3511337E-02 | 5.2985449E-02 | 5.2140980E-02 | 4.9686190E-02 | 3.6854018E-02 | 2.0230769E-02 | 0.0000000E+00 |
| -3.000E-01 | 6.1542012E-02 | 6.1991418E-02 | 6.1978625E-02 | 6.0886294E-02 | 4.9616521E-02 | 2.9827680E-02 | 0.0000000E+00 |
| -2.000E-01 | 6.6956243E-02 | 6.9056402E-02 | 7.0499635E-02 | 7.2037240E-02 | 6.6696988E-02 | 4.6300639E-02 | 0.0000000E+00 |
| -1.000E-01 | 6.5529583E-02 | 7.0004105E-02 | 7.3432378E-02 | 7.8590386E-02 | 8.4462845E-02 | 7.3611021E-02 | 0.0000000E+00 |
| 0.000E+00 | 5.1748534E-02 | 6.1709609E-02 | 6.8016336E-02 | 7.7466538E-02 | 9.3997859E-02 | 9.7483668E-02 | 0.0000000E+00 |
| 0.000E+00 | 0.0000000E+00 | 6.1709609E-02 | 6.8016336E-02 | 7.7466538E-02 | 9.3997859E-02 | 9.7483668E-02 | 7.9312594E-02 |
| 1.000E-01 | 0.0000000E+00 | 2.2494931E-02 | 3.9518122E-02 | 6.2652973E-02 | 9.6254337E-02 | 1.0871841E-01 | 1.0818930E-01 |
| 2.000E-01 | 0.0000000E+00 | 1.3100677E-02 | 2.5340076E-02 | 4.6772905E-02 | 9.1591615E-02 | 1.1381468E-01 | 1.2421167E-01 |
| 3.000E-01 | 0.0000000E+00 | 1.0194331E-02 | 2.0270341E-02 | 3.9472225E-02 | 8.7467481E-02 | 1.1662007E-01 | 1.3571209E-01 |
| 4.000E-01 | 0.0000000E+00 | 9.5290644E-03 | 1.9067650E-02 | 3.7770256E-02 | 8.8332315E-02 | 1.2250259E-01 | 1.4826767E-01 |
| 5.000E-01 | 0.0000000E+00 | 1.0263750E-02 | 2.0502330E-02 | 4.0649220E-02 | 9.6418345E-02 | 1.3581653E-01 | 1.6751584E-01 |
| 6.000E-01 | 0.0000000E+00 | 1.2529327E-02 | 2.4909477E-02 | 4.9065634E-02 | 1.1533634E-01 | 1.6223048E-01 | 2.0070062E-01 |
| 7.000E-01 | 0.0000000E+00 | 1.7417124E-02 | 3.4415206E-02 | 6.7081148E-02 | 1.5398186E-01 | 2.1356335E-01 | 2.6167192E-01 |
| 8.000E-01 | 0.0000000E+00 | 2.8562211E-02 | 5.6020429E-02 | 1.0769702E-01 | 2.3848565E-01 | 3.2254956E-01 | 3.8692070E-01 |
| 9.000E-01 | 0.0000000E+00 | 6.1633112E-02 | 1.1976311E-01 | 2.2612375E-01 | 4.7610276E-01 | 6.1970346E-01 | 7.1774509E-01 |
| 1.000E+00 | 0.0000000E+00 | 3.2812354E-01 | 6.2906510E-01 | 1.1563161E+00 | 2.2483946E+00 | 2.7414726E+00 | 2.9776602E+00 |

Table 6a
Scattering Coefficients for Cloud $C_1$ phase function

| l | $\beta_l$ | $\beta_{l+35}$ | $\beta_{l+70}$ | $\beta_{l+105}$ | $\beta_{l+140}$ | $\beta_{l+175}$ | $\beta_{l+210}$ | $\beta_{l+245}$ | $\beta_{l+280}$ |
|---|---|---|---|---|---|---|---|---|---|
| 0 | 1.000 | 19.884 | 16.144 | 6.990 | 2.025 | 0.440 | 0.079 | 0.012 | 0.002 |
| 1 | 2.544 | 20.024 | 15.883 | 6.785 | 1.940 | 0.422 | 0.074 | 0.011 | 0.002 |
| 2 | 3.883 | 20.145 | 15.606 | 6.573 | 1.869 | 0.401 | 0.071 | 0.011 | 0.001 |
| 3 | 4.568 | 20.251 | 15.338 | 6.377 | 1.790 | 0.384 | 0.067 | 0.010 | 0.001 |
| 4 | 5.235 | 20.330 | 15.058 | 6.173 | 1.723 | 0.364 | 0.064 | 0.009 | 0.001 |
| 5 | 5.887 | 20.401 | 14.784 | 5.986 | 1.649 | 0.349 | 0.060 | 0.009 | 0.001 |
| 6 | 6.457 | 20.444 | 14.501 | 5.790 | 1.588 | 0.331 | 0.057 | 0.008 | 0.001 |
| 7 | 7.177 | 20.477 | 14.225 | 5.612 | 1.518 | 0.317 | 0.054 | 0.008 | 0.001 |
| 8 | 7.859 | 20.489 | 13.941 | 5.424 | 1.461 | 0.301 | 0.052 | 0.008 | 0.001 |
| 9 | 8.494 | 20.483 | 13.662 | 5.255 | 1.397 | 0.288 | 0.049 | 0.007 | 0.001 |
| 10 | 9.286 | 20.467 | 13.378 | 5.075 | 1.344 | 0.273 | 0.047 | 0.007 | 0.001 |
| 11 | 9.856 | 20.427 | 13.098 | 4.915 | 1.284 | 0.262 | 0.044 | 0.006 | 0.001 |
| 12 | 10.615 | 20.382 | 12.816 | 4.744 | 1.235 | 0.248 | 0.042 | 0.006 | 0.001 |
| 13 | 11.229 | 20.310 | 12.536 | 4.592 | 1.179 | 0.238 | 0.039 | 0.006 | 0.001 |
| 14 | 11.851 | 20.236 | 12.257 | 4.429 | 1.134 | 0.225 | 0.038 | 0.005 | 0.001 |
| 15 | 12.503 | 20.136 | 11.978 | 4.285 | 1.082 | 0.215 | 0.035 | 0.005 | 0.001 |
| 16 | 13.058 | 20.036 | 11.703 | 4.130 | 1.040 | 0.204 | 0.034 | 0.005 | 0.001 |
| 17 | 13.626 | 19.909 | 11.427 | 3.994 | 0.992 | 0.195 | 0.032 | 0.005 | 0.001 |
| 18 | 14.209 | 19.785 | 11.156 | 3.847 | 0.954 | 0.185 | 0.030 | 0.004 | 0.001 |
| 19 | 14.660 | 19.632 | 10.884 | 3.719 | 0.909 | 0.177 | 0.029 | 0.004 | 0.001 |
| 20 | 15.231 | 19.486 | 10.618 | 3.580 | 0.873 | 0.167 | 0.027 | 0.004 | |
| 21 | 15.641 | 19.311 | 10.350 | 3.459 | 0.832 | 0.160 | 0.026 | 0.004 | |
| 22 | 16.126 | 19.145 | 10.090 | 3.327 | 0.799 | 0.151 | 0.024 | 0.003 | |
| 23 | 16.539 | 18.949 | 9.827 | 3.214 | 0.762 | 0.145 | 0.023 | 0.003 | |
| 24 | 16.934 | 18.764 | 9.574 | 3.090 | 0.731 | 0.137 | 0.022 | 0.003 | |
| 25 | 17.325 | 18.551 | 9.318 | 2.983 | 0.696 | 0.131 | 0.021 | 0.003 | |
| 26 | 17.673 | 18.348 | 9.072 | 2.866 | 0.668 | 0.124 | 0.020 | 0.003 | |
| 27 | 17.999 | 18.119 | 8.822 | 2.766 | 0.636 | 0.118 | 0.018 | 0.003 | |
| 28 | 18.329 | 17.901 | 8.584 | 2.656 | 0.610 | 0.112 | 0.018 | 0.002 | |
| 29 | 18.588 | 17.659 | 8.340 | 2.562 | 0.581 | 0.107 | 0.017 | 0.002 | |
| 30 | 18.885 | 17.428 | 8.110 | 2.459 | 0.557 | 0.101 | 0.016 | 0.002 | |
| 31 | 19.103 | 17.174 | 7.874 | 2.372 | 0.530 | 0.097 | 0.015 | 0.002 | |
| 32 | 19.345 | 16.931 | 7.652 | 2.274 | 0.508 | 0.091 | 0.014 | 0.002 | |
| 33 | 19.537 | 16.668 | 7.424 | 2.193 | 0.483 | 0.087 | 0.013 | 0.002 | |
| 34 | 19.721 | 16.415 | 7.211 | 2.102 | 0.463 | 0.082 | 0.013 | 0.002 | |

Table 6b

Angular Intensity for Cloud $C_1$ Scattering Kernel ($L = 300$)

($\omega = 1$, $\mu_0 = 1$, $\tau_0 = 64$, $I_{inc} = 0.5$)

| $\mu \backslash \tau$ | 0 | $\tau_0/20$ | $\tau_0/10$ | $\tau_0/5$ | $\tau_0/2$ | $3\tau_0/4$ | $\tau_0$ |
|---|---|---|---|---|---|---|---|
| -1.000E+00 | 1.0636984E+00 | 1.0062387E+00 | 9.6320640E-01 | 8.5824229E-01 | 5.2453336E-01 | 2.4600228E-01 | 0.0000000E+00 |
| -9.000E-01 | 9.5309008E-01 | 9.9566229E-01 | 9.6972419E-01 | 8.6938979E-01 | 5.3598880E-01 | 2.5740482E-01 | 0.0000000E+00 |
| -8.000E-01 | 9.5407647E-01 | 9.9828274E-01 | 9.7776589E-01 | 8.8052819E-01 | 5.4744427E-01 | 2.6883574E-01 | 0.0000000E+00 |
| -7.000E-01 | 8.8254184E-01 | 9.8850614E-01 | 9.8351863E-01 | 8.9156832E-01 | 5.5889973E-01 | 2.8028148E-01 | 0.0000000E+00 |
| -6.000E-01 | 8.2471232E-01 | 9.7909867E-01 | 9.8890358E-01 | 9.0255626E-01 | 5.7035518E-01 | 2.9173427E-01 | 0.0000000E+00 |
| -5.000E-01 | 7.7260568E-01 | 9.6977241E-01 | 9.9399054E-01 | 9.1349749E-01 | 5.8181061E-01 | 3.0319021E-01 | 0.0000000E+00 |
| -4.000E-01 | 7.1143850E-01 | 9.5800446E-01 | 9.9832954E-01 | 9.2437584E-01 | 5.9326601E-01 | 3.1464747E-01 | 0.0000000E+00 |
| -3.000E-01 | 6.4031056E-01 | 9.4342529E-01 | 1.0017757E+00 | 9.3517893E-01 | 6.0472139E-01 | 3.2610518E-01 | 0.0000000E+00 |
| -2.000E-01 | 5.5848173E-01 | 9.2583435E-01 | 1.0042140E+00 | 9.4589447E-01 | 6.1617674E-01 | 3.3756297E-01 | 0.0000000E+00 |
| -1.000E-01 | 4.5873404E-01 | 9.0459251E-01 | 1.0054770E+00 | 9.5650944E-01 | 6.2763204E-01 | 3.4902064E-01 | 0.0000000E+00 |
| 0.000E+00 | 2.5158245E-01 | 8.7951999E-01 | 1.0054752E+00 | 9.6701431E-01 | 6.3908730E-01 | 3.6047812E-01 | 0.0000000E+00 |
| 0.000E+00 | 0.0000000E+00 | 8.7951999E-01 | 1.0054752E+00 | 9.6701431E-01 | 6.3908730E-01 | 3.6047812E-01 | 3.9263859E-02 |
| 1.000E-01 | 0.0000000E+00 | 8.5069625E-01 | 1.0042130E+00 | 9.7740715E-01 | 6.5054252E-01 | 3.7193540E-01 | 7.2039069E-02 |
| 2.000E-01 | 0.0000000E+00 | 8.1871840E-01 | 1.0018813E+00 | 9.8770148E-01 | 6.6199770E-01 | 3.8339248E-01 | 8.8948974E-02 |
| 3.000E-01 | 0.0000000E+00 | 7.8521516E-01 | 9.9901551E-01 | 9.9794090E-01 | 6.7345285E-01 | 3.9484937E-01 | 1.0414040E-01 |
| 4.000E-01 | 0.0000000E+00 | 7.5459856E-01 | 9.9678835E-01 | 1.0082266E+00 | 6.8490802E-01 | 4.0630611E-01 | 1.1838392E-01 |
| 5.000E-01 | 0.0000000E+00 | 7.3516216E-01 | 9.9760279E-01 | 1.0187710E+00 | 6.9636329E-01 | 4.1776272E-01 | 1.3199043E-01 |
| 6.000E-01 | 0.0000000E+00 | 7.3765495E-01 | 1.0059570E+00 | 1.0300017E+00 | 7.0781884E-01 | 4.2921921E-01 | 1.4513931E-01 |
| 7.000E-01 | 0.0000000E+00 | 7.7792408E-01 | 1.0300617E+00 | 1.0427819E+00 | 7.1927503E-01 | 4.4067560E-01 | 1.5795774E-01 |
| 8.000E-01 | 0.0000000E+00 | 8.8704554E-01 | 1.0859454E+00 | 1.0589106E+00 | 7.3073267E-01 | 4.5213189E-01 | 1.7054334E-01 |
| 9.000E-01 | 0.0000000E+00 | 1.1539390E+00 | 1.2141267E+00 | 1.0826504E+00 | 7.4219359E-01 | 4.6358809E-01 | 1.8295432E-01 |
| 1.000E+00 | 0.0000000E+00 | 8.0745963E+01 | 1.1786122E+01 | 1.2605960E+00 | 7.5366350E-01 | 4.7504419E-01 | 1.9523120E-01 |

Table 6c

Angular Intensity for Cloud $C_1$ Scattering Kernel ($L = 300$)

($\omega = 0.9$, $\mu_0 = 1$, $\tau_0 = 64$, $I_{inc} = 0.5$)

| $\mu \backslash \tau$ | 0 | $\tau_0/20$ | $\tau_0/10$ | $\tau_0/5$ | $\tau_0/2$ | $3\tau_0/4$ | $\tau_0$ |
|---|---|---|---|---|---|---|---|
| -1.000E+00 | 2.0977263E-01 | 8.6612558E-02 | 4.1343507E-02 | 9.5110502E-03 | 1.0826714E-04 | 2.5785810E-06 | 0.0000000E+00 |
| -9.000E-01 | 1.3305687E-01 | 7.9916028E-02 | 4.1642268E-02 | 9.8885842E-03 | 1.1300385E-04 | 2.6919500E-06 | 0.0000000E+00 |
| -8.000E-01 | 1.5585660E-01 | 8.4979470E-02 | 4.3751627E-02 | 1.0415194E-02 | 1.1926997E-04 | 2.8417201E-06 | 0.0000000E+00 |
| -7.000E-01 | 1.2247674E-01 | 8.3044807E-02 | 4.5314646E-02 | 1.1071380E-02 | 1.2734599E-04 | 3.0345968E-06 | 0.0000000E+00 |
| -6.000E-01 | 1.0613103E-01 | 8.3816646E-02 | 4.7705518E-02 | 1.1909143E-02 | 1.3756086E-04 | 3.2784250E-06 | 0.0000000E+00 |
| -5.000E-01 | 1.0037246E-01 | 8.7262657E-02 | 5.1102514E-02 | 1.2965758E-02 | 1.5030235E-04 | 3.5824565E-06 | 0.0000000E+00 |
| -4.000E-01 | 9.3636118E-02 | 9.1881918E-02 | 5.5402754E-02 | 1.4275454E-02 | 1.6603111E-04 | 3.9576828E-06 | 0.0000000E+00 |
| -3.000E-01 | 8.6109828E-02 | 9.7787475E-02 | 6.0730179E-02 | 1.5882618E-02 | 1.8529814E-04 | 4.4172513E-06 | 0.0000000E+00 |
| -2.000E-01 | 7.8467545E-02 | 1.0521444E-01 | 6.7252374E-02 | 1.7841070E-02 | 2.0876660E-04 | 4.9769831E-06 | 0.0000000E+00 |
| -1.000E-01 | 6.7611157E-02 | 1.1404115E-01 | 7.5124637E-02 | 2.0215257E-02 | 2.3723909E-04 | 5.6560242E-06 | 0.0000000E+00 |
| 0.000E+00 | 4.0639285E-02 | 1.2442938E-01 | 8.4570625E-02 | 2.3084120E-02 | 2.7169213E-04 | 6.4776672E-06 | 0.0000000E+00 |
| 0.000E+00 | 0.0000000E+00 | 1.2442938E-01 | 8.4570625E-02 | 2.3084120E-02 | 2.7169213E-04 | 6.4776672E-06 | 7.6444566E-08 |
| 1.000E-01 | 0.0000000E+00 | 1.3659785E-01 | 9.5875146E-02 | 2.6544808E-02 | 3.1332012E-04 | 7.4704005E-06 | 1.3641808E-07 |
| 2.000E-01 | 0.0000000E+00 | 1.5094716E-01 | 1.0941827E-01 | 3.0718008E-02 | 3.6359187E-04 | 8.6692542E-06 | 1.7437041E-07 |
| 3.000E-01 | 0.0000000E+00 | 1.6826506E-01 | 1.2573334E-01 | 3.5755721E-02 | 4.2432391E-04 | 1.0117547E-05 | 2.1525686E-07 |
| 4.000E-01 | 0.0000000E+00 | 1.9021700E-01 | 1.4562207E-01 | 4.1853393E-02 | 4.9777650E-04 | 1.1869170E-05 | 2.6167378E-07 |
| 5.000E-01 | 0.0000000E+00 | 2.2019044E-01 | 1.7041619E-01 | 4.9270986E-02 | 5.8678087E-04 | 1.3991604E-05 | 3.1578237E-07 |
| 6.000E-01 | 0.0000000E+00 | 2.6371238E-01 | 2.0243123E-01 | 5.8372517E-02 | 6.9490973E-04 | 1.6569950E-05 | 3.7993345E-07 |
| 7.000E-01 | 0.0000000E+00 | 3.3126941E-01 | 2.4609184E-01 | 6.9714335E-02 | 8.2671008E-04 | 1.9712361E-05 | 4.5692954E-07 |
| 8.000E-01 | 0.0000000E+00 | 4.4603725E-01 | 3.1086587E-01 | 8.4268691E-02 | 9.8802957E-04 | 2.3557470E-05 | 5.5024338E-07 |
| 9.000E-01 | 0.0000000E+00 | 6.7671565E-01 | 4.2268339E-01 | 1.0417393E-01 | 1.1865029E-03 | 2.8284673E-05 | 6.6425243E-07 |
| 1.000E+00 | 0.0000000E+00 | 6.9479416E+01 | 8.9756634E+00 | 2.2308547E-01 | 1.4326865E-03 | 3.4128613E-05 | 8.0461843E-07 |

Table 7
Angular Quadrature Order at Convergence for Tables 5b,c and 6b,c

| Kernel | $\omega$ | 2N(5-place) | 2N(7-place) | Convergence Mode |
|---|---|---|---|---|
| HAZE-L | 1 | 72 | 552 | Original |
| HAZE-L | 0.9 | 72 | 100 | Original |
| Cloud-$C_1$ | 1 | 288 | 356 | Original |
| Cloud-$C_1$ | 0.9 | 304 | 372 | Original |

In summary, the RM/DOM algorithm apparently has no difficulty in reproducing the highest quality benchmarks found in the literature for angular intensity as suggested by the Radiation Commission and presented by Garcia and Siewert [7] for beam incidence. In addition, we have extend the benchmarks two additional digits for greater impact. Transmittance and reflectance are also of 7- place accuracy. It is interesting to note that convergence acceleration is most effective for integral quantities. In general, experience shows that convergence acceleration is most effective for thin slabs, but also depends on the phase function representation, desired error and the initial order $N_0$ of the sequence. Including convergence acceleration provides additional confirmation of the benchmark results as presented here. Finally, the algorithm is equally applicable for Fourier components other than zero, at least for low order scattering, a feature expected to hold for high order scattering as well.

As we now show, the RM/DOM algorithm is particularly convenient for heterogeneous media.

**IV. Numerical Demonstration for a Heterogeneous Medium**
The form of Eqs(32) suggests the analysis to follow for a heterogeneous medium.

Consider a composite medium made of $l$-1 homogeneous slabs together with a single homogeneous slab as shown in Fig. 2. Let the composite response matrix $Q_{l-1}$ for the composite slab as inferred from Eq(32a) be defined by the action

$$\begin{bmatrix} I_0^- \\ I_{l-1}^+ \end{bmatrix} = Q_{l-1} \begin{bmatrix} I_{l-1}^- \\ I_0^+ \end{bmatrix} + \begin{bmatrix} s_{l-1,1} \\ s_{l-1,2} \end{bmatrix} \qquad (54a)$$

including composite sources, where the ingoing/outgoing angular intensity at surface $l$ is

$$I_l^\pm \equiv I^\pm(\tau_l). \tag{54b}$$

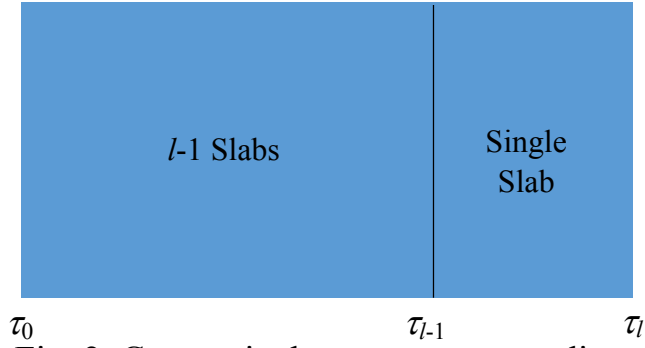

Fig. 2. Composite heterogeneous medium.

Similarly, we now write the single slab response from Eq(32b)

$$\begin{bmatrix} I_{l-1}^- \\ I_l^+ \end{bmatrix} = R_l \begin{bmatrix} I_l^- \\ I_{l-1}^+ \end{bmatrix} + \begin{bmatrix} \tilde{s}_{l,1} \\ \tilde{s}_{l,2} \end{bmatrix}. \tag{55}$$

If, we partition $Q_l$ and $R_l$ into four $N$ by $N$ matrices

$$Q_l = \begin{bmatrix} Q_{l,1} & Q_{l,2} \\ Q_{l,3} & Q_{l,4} \end{bmatrix}$$

$$R_l = \begin{bmatrix} T_{l,n} & R_{l,f} \\ R_{l,f} & T_{l,n} \end{bmatrix} \tag{56a,b}$$

Eqs(54a) and (55) resolve into

$$\begin{aligned}
I_0^- &= Q_{l-1,1}I_{l-1}^- + Q_{l-1,2}I_0^+ + s_{l-1,1} \\
I_l^+ &= R_{l,f}I_l^- + T_{l,n}I_{l-1}^+ + \tilde{s}_{l,2} \\
I_{l-1}^+ &= Q_{l-1,3}I_{l-1}^- + Q_{l-1,4}I_0^+ + s_{l-1,2} \\
I_{l-1}^- &= T_{l,n}I_l^- + R_{l,f}I_{l-1}^+ + \tilde{s}_{l,1}.
\end{aligned}$$

(56b)

On combining the second and third, and first and fourth equations, we find

$$\begin{bmatrix} I & -Q_{l-1,3} \\ -R_{l,f} & I \end{bmatrix} \begin{bmatrix} I_{l-1}^+ \\ I_{l-1}^- \end{bmatrix} = \begin{bmatrix} 0 & -Q_{l-1,4} \\ -T_{l,n} & 0 \end{bmatrix} \begin{bmatrix} I_l^- \\ I_0^+ \end{bmatrix} + \begin{bmatrix} s_{l-1,2} \\ \tilde{s}_{l,1} \end{bmatrix}$$

(57a)

$$\begin{bmatrix} I_0^- \\ I_l^+ \end{bmatrix} = \begin{bmatrix} 0 & Q_{l-1,2} \\ R_{l,f} & 0 \end{bmatrix} \begin{bmatrix} I_l^- \\ I_0^+ \end{bmatrix} + \begin{bmatrix} 0 & Q_{l-1,1} \\ T_{l,n} & 0 \end{bmatrix} \begin{bmatrix} I_{l-1}^+ \\ I_{l-1}^- \end{bmatrix} + \begin{bmatrix} s_{l-1,1} \\ \tilde{s}_{l,2} \end{bmatrix}.$$

(57b)

Now solve for the intensities at the interface from Eq(57a) to give

$$\begin{bmatrix} I_{l-1}^+ \\ I_{l-1}^- \end{bmatrix} = U_l \begin{bmatrix} I_l^- \\ I_0^+ \end{bmatrix} + V_l \begin{bmatrix} s_{l-1,2} \\ \tilde{s}_{l,1} \end{bmatrix},$$

(58a)

where

$$w_{l-} \equiv \begin{bmatrix} I_N & -Q_{l-1,3} \\ -R_{l,f} & I_N \end{bmatrix}$$

$$w_{l+} \equiv \begin{bmatrix} 0 & -Q_{l-1,4} \\ -T_{l,n} & 0 \end{bmatrix}.$$

(58b,c)

$$V_l \equiv w_{l-}^{-1}$$
$$U_l \equiv V_l w_{l+}.$$

(58d,e)

Introducing the result into Eq(57b) then gives

$$\begin{bmatrix} I_0^- \\ I_l^+ \end{bmatrix} = \left\{ \begin{bmatrix} 0 & Q_{l-1,1} \\ T_{l,n} & 0 \end{bmatrix} U_l + \begin{bmatrix} 0 & Q_{l-1,2} \\ R_{l,f} & 0 \end{bmatrix} \right\} \begin{bmatrix} I_l^- \\ I_0^+ \end{bmatrix} +$$
$$+ \begin{bmatrix} 0 & Q_{l-1,1} \\ T_{l,n} & 0 \end{bmatrix} V_l \begin{bmatrix} s_{l-1,2} \\ \tilde{s}_{l,1} \end{bmatrix} + \begin{bmatrix} s_{l-1,1} \\ \tilde{s}_{l,2} \end{bmatrix}. \tag{59}$$

From the definition of the composite response $Q_l$ and source $s_l$ in Eq(54a) therefore,

$$Q_l = \begin{bmatrix} Q_{l,1} & Q_{l,2} \\ Q_{l,3} & Q_{l,4} \end{bmatrix} = \begin{bmatrix} 0 & Q_{l-1,1} \\ T_{l,n} & 0 \end{bmatrix} U_l + \begin{bmatrix} 0 & Q_{l-1,2} \\ R_{l,f} & 0 \end{bmatrix} \tag{60a}$$

$$\begin{bmatrix} s_{l,1} \\ s_{l,2} \end{bmatrix} \equiv \begin{bmatrix} 0 & Q_{l-1,1} \\ T_{l,n} & 0 \end{bmatrix} V_l \begin{bmatrix} s_{l-1,2} \\ \tilde{s}_{l,1} \end{bmatrix} + \begin{bmatrix} s_{l-1,1} \\ \tilde{s}_{l,2} \end{bmatrix} \tag{60b}$$

to give the exiting intensity from the composite $l$-1 plus homogeneous slabs

$$\begin{bmatrix} I_0^- \\ I_l^+ \end{bmatrix} = Q_l \begin{bmatrix} I_l^- \\ I_0^+ \end{bmatrix} + \begin{bmatrix} s_{l,1} \\ s_{l,2} \end{bmatrix}. \tag{60c}$$

Each of the four $N$-by-$N$ components of the matrix $Q_l$ and the two $N$-by-$N$ components of the vector $s_l$ are therefore recursively obtained.

The recurrence begins with the degenerate case of a composite of $l = 1$

$$\begin{bmatrix} I_0^- \\ I_0^+ \end{bmatrix} = Q_0 \begin{bmatrix} I_0^- \\ I_0^+ \end{bmatrix} + \begin{bmatrix} s_{0,1} \\ s_{0,2} \end{bmatrix} \tag{61a}$$

resulting in (for consistency)

$$Q_0 = I_{2N}$$
$$\begin{bmatrix} s_{0,1} \\ s_{0,2} \end{bmatrix} = 0. \tag{61b}$$

As a check, let $l = 1$, then from Eqs(58b,c)

$$w_{1-} \equiv \begin{bmatrix} I_N & 0 \\ -R_{1,f} & I_N \end{bmatrix}$$

$$w_{1+} \equiv \begin{bmatrix} 0 & -I_N \\ -T_{1,n} & 0 \end{bmatrix},$$

which gives for $w_{1-}^{-1}$

$$w_{1-}^{-1T} \equiv \begin{bmatrix} I_N & -R_{1,f} \\ 0 & I_N \end{bmatrix}^{-1} \Rightarrow w_{1-}^{-1} \equiv \begin{bmatrix} I_N & 0 \\ R_{1,f} & I_N \end{bmatrix}$$

and therefore

$$w_{1-}^{-1} w_{1+} \equiv \begin{bmatrix} 0 & I_N \\ T_{1,n} & R_{1,f} \end{bmatrix}.$$

When introduced into Eq(58d,e), Eq(60a) becomes

$$Q_1 = \begin{bmatrix} 0 & I_N \\ T_{1,n} & 0 \end{bmatrix} \begin{bmatrix} 0 & I_N \\ T_{1,n} & R_{1,f} \end{bmatrix} + \begin{bmatrix} 0 & 0 \\ R_{1,f} & 0 \end{bmatrix} = \begin{bmatrix} T_{1,n} & R_{1,f} \\ R_{1,f} & T_{1,n} \end{bmatrix},$$

which is $R_1$ as required. Similarly, one can show

$$\begin{bmatrix} s_{1,1} \\ s_{1,2} \end{bmatrix} = \begin{bmatrix} \tilde{s}_{1,1} \\ \tilde{s}_{1,2} \end{bmatrix}.$$

Now consider a heterogeneous medium composed of $n$ slabs with entering angular intensities $I_0^+$ and $I_n^-$ at the free surfaces. Since one knows the composite $n$-slab response and source from Eqs(60a,b), one knows the exiting exterior surface angular intensities from Eq(60c) as well

$$\begin{bmatrix} I_0^- \\ I_n^+ \end{bmatrix} = Q_n \begin{bmatrix} I_n^- \\ I_0^+ \end{bmatrix} + \begin{bmatrix} s_{n,1} \\ s_{n,2} \end{bmatrix}. \tag{62a}$$

The interior slab interfacial intensities then immediately follow from Eq(58a) as a recurrence. Since $I_n^-$ and $I_0^+$ are both known, by expanding the partitioned matrix multiplication, there results a backward recurrence for $l = n, n\text{-}1,\ldots,2$

$$\begin{aligned} I_{l-1}^- &= U_{l,3} I_l^- + U_{l,4} I_0^+ + V_{l,3} s_{l-1,2} + V_{l,4} \tilde{s}_{l,1} \\ I_{l-1}^+ &= U_{l,1} I_l^- + U_{l,2} I_0^+ + V_{l,1} s_{l-1,2} + V_{l,2} \tilde{s}_{l,1} \end{aligned} \tag{62b}$$

giving the intensities at interface $l$-1 and continuing to $l = 2$.

As a demonstration, consider a heterogeneous medium of 10 slabs with the properties in Table 8a. Here, we assume a Henyey-Greenstein (H-G) scattering phase function for all slabs but with a gradient of asymmetry factors from top to bottom of the medium. A normalized perpendicular beam shines on the top surface, and the medium is conservative. The medium simulates a cloud with a vertically decreasing aerosol gradient.

Table 8a
Heterogeneous slab optical properties

| $\omega$ | $\tau$ | $g$ |
|---|---|---|
| 1 | 200.0 | 0.95 |
| 1 | 100.0 | 0.85 |
| 1 | 50.0 | 0.75 |
| 1 | 25.0 | 0.65 |
| 1 | 12.5 | 0.55 |
| 1 | 6.25 | 0.45 |
| 1 | 3.125 | 0.35 |
| 1 | 1.5625 | 0.25 |
| 1 | 0.78125 | 0.15 |
| 1 | 0.390125 | 0.05 |

To make this benchmark more interesting, we use the exact closed form H-G kernel [27] expressed in terms of the elliptic integral $E(k)$

$$f(\mu',\mu) = \frac{2}{\pi} \frac{1-g^2(\tau)}{(a+b)^{3/2}} \frac{E\left(\sqrt{\frac{2b}{a+b}}\right)}{1-\frac{2b}{a+b}} \tag{63a}$$

where

$$a = 1 + g^2(\tau) - 2g(\tau) - \mu\mu'$$
$$b = 2g(\tau)\sqrt{1-\mu^2}\sqrt{1-\mu'^2}$$
$$E(k) \equiv \int_0^{\pi/2} d\theta \sqrt{1-k^2 \sin^2(\theta)},$$

rather than the usual Legendre expansion for which

$$\omega_l(\tau) \equiv (2l+1)g^l(\tau). \tag{63b}$$

Table 8 gives the interfacial angular intensities to 7-places. For this benchmark, all slabs converged by *W-e*. Note that nearly the same table of values (to one unit in the last place) was calculated using the Legendre representation of the scattering kernel. In addition, a flux calculation verified the flux to be spatially uniform to 8-places. Table 8b gives the final verification— the convergence of $R_f+T_n$ to unity. We now see the power of convergence acceleration. At $N = 102$, *W-e* converged to 1 to within one unit in ninth place; whereas, the original sequence has converged to only 3-places. *W-e* remains converged until $N = 146$ (when the calculation stopped) at which time the original sequence finally converged to an acceptable 6- places. There is a savings of nearly 52 quadrature points (94 vs 146) with *W-e*.

The analysis presented above is actually an analytical update of the star product formulation [10] which leads to a formal theory of radiative based on the principles of invariance. Here, the method is in the form of matrices, which generalize to operators leading to a solution continuous in all variables.

The advantage of the analytical star product formulation as derived is that we know the response matrix for each slab from Eqs(33) avoiding the doubling procedure or any other determination of the response matrix such as through Chandrasekhar's X and Y functions that usually accompanies the star product formalism.

**Conclusion**

In conclusion, there is a new solution to the azimuthally symmetric RTE called the response matrix DOM (RM/DOM). By assuming eigenvalues and eigenvectors, the solution features an alternative representation that naturally avoids the inherent instability in common exponential solutions without the need for an ad-hoc fix. An explicit solution representation results in the form of matrix hyperbolic functions, eliminating the matrix inversion required for the coefficients of an exponentially based representation. A "faux quadrature" permits the RM/DOM solution itself to generate the necessary interpolation for angular edits by adding the edits with zero weight to the quadrature list. Convergence acceleration, through a Wynn-epsilon algorithm, enables extreme accuracy of 8- places for reflectance and transmittance and 7- places for the angular intensity. Confirmation of the claimed accuracy is through published benchmarks and internal particle conservation. While, in general, 6- place accuracy is certainly more than sufficient for any application, benchmarking requires extreme accuracy, which, as we see, is attainable by the RM/DOM algorithm.

The primary importance of the RM/DOM solution is not just that it provides accurate results for the RTE, but that it does so in a mathematically consistent and efficient way that leads to a more transparent numerical evaluation than found previously. Possible extensions include multi- dimensions since the solution is essentially a nodal method, which lends itself to transverse integration to treat a second dimension.

Table 8b

($\omega = 1$, $\mu_0 = 1$, $\tau_0 = 100$, $I_{inc} = 0.5$, $N = 200$)

Slabs 1-5

| $\mu$\Interface | Top | 1 | 2 | 3 | 4 | 5 |
|---|---|---|---|---|---|---|
| -1.000E+00 | 7.7952790E-01 | 4.3673798E-01 | 3.0375049E-01 | 1.9292759E-01 | 1.1535156E-01 | 6.5481258E-02 |
| -9.000E-01 | 8.1635174E-01 | 4.3762456E-01 | 3.0463708E-01 | 1.9381417E-01 | 1.1623814E-01 | 6.6367831E-02 |
| -8.000E-01 | 8.6213715E-01 | 4.3851115E-01 | 3.0552366E-01 | 1.9470076E-01 | 1.1712472E-01 | 6.7254413E-02 |
| -7.000E-01 | 9.1984402E-01 | 4.3939773E-01 | 3.0641024E-01 | 1.9558734E-01 | 1.1801131E-01 | 6.8140997E-02 |
| -6.000E-01 | 9.9335714E-01 | 4.4028431E-01 | 3.0729683E-01 | 1.9647392E-01 | 1.1889789E-01 | 6.9027580E-02 |
| -5.000E-01 | 1.0873289E+00 | 4.4117090E-01 | 3.0818341E-01 | 1.9736050E-01 | 1.1978447E-01 | 6.9914164E-02 |
| -4.000E-01 | 1.2058839E+00 | 4.4205748E-01 | 3.0906999E-01 | 1.9824709E-01 | 1.2067105E-01 | 7.0800747E-02 |
| -3.000E-01 | 1.3476718E+00 | 4.4294406E-01 | 3.0995658E-01 | 1.9913367E-01 | 1.2155764E-01 | 7.1687330E-02 |
| -2.000E-01 | 1.4895513E+00 | 4.4383065E-01 | 3.1084316E-01 | 2.0002025E-01 | 1.2244422E-01 | 7.2573914E-02 |
| -1.000E-01 | 1.5307094E+00 | 4.4471723E-01 | 3.1172974E-01 | 2.0090684E-01 | 1.2333080E-01 | 7.3460497E-02 |
| 0.000E+00 | 8.1079477E-01 | 4.4560381E-01 | 3.1261632E-01 | 2.0179342E-01 | 1.2421739E-01 | 7.4347080E-02 |
| 0.000E+00 | 0.0000000E+00 | 4.4560381E-01 | 3.1261632E-01 | 2.0179342E-01 | 1.2421739E-01 | 7.4347080E-02 |
| 1.000E-01 | 0.0000000E+00 | 4.4649039E-01 | 3.1350291E-01 | 2.0268000E-01 | 1.2510397E-01 | 7.5233664E-02 |
| 2.000E-01 | 0.0000000E+00 | 4.4737698E-01 | 3.1438949E-01 | 2.0356659E-01 | 1.2599055E-01 | 7.6120247E-02 |
| 3.000E-01 | 0.0000000E+00 | 4.4826356E-01 | 3.1527607E-01 | 2.0445317E-01 | 1.2687714E-01 | 7.7006830E-02 |
| 4.000E-01 | 0.0000000E+00 | 4.4915014E-01 | 3.1616266E-01 | 2.0533975E-01 | 1.2776372E-01 | 7.7893413E-02 |
| 5.000E-01 | 0.0000000E+00 | 4.5003673E-01 | 3.1704924E-01 | 2.0622634E-01 | 1.2865030E-01 | 7.8779997E-02 |
| 6.000E-01 | 0.0000000E+00 | 4.5092331E-01 | 3.1793582E-01 | 2.0711292E-01 | 1.2953689E-01 | 7.9666580E-02 |
| 7.000E-01 | 0.0000000E+00 | 4.5180989E-01 | 3.1882241E-01 | 2.0799950E-01 | 1.3042347E-01 | 8.0553163E-02 |
| 8.000E-01 | 0.0000000E+00 | 4.5269648E-01 | 3.1970899E-01 | 2.0888609E-01 | 1.3131005E-01 | 8.1439746E-02 |
| 9.000E-01 | 0.0000000E+00 | 4.5358306E-01 | 3.2059557E-01 | 2.0977267E-01 | 1.3219664E-01 | 8.2326330E-02 |
| 1.000E+00 | 0.0000000E+00 | 4.5446964E-01 | 3.2148216E-01 | 2.1065925E-01 | 1.3308322E-01 | 8.3212913E-02 |

Slabs 6-10

| $\mu$\Interface | 6 | 7 | 8 | 9 | Bottom |
|---|---|---|---|---|---|
| -1.000E+00 | 3.5009728E-02 | 1.7113552E-02 | 7.2362544E-03 | 2.2705529E-03 | 0.0000000E+00 |
| -9.000E-01 | 3.5892001E-02 | 1.7916412E-02 | 7.7779228E-03 | 2.4925953E-03 | 0.0000000E+00 |
| -8.000E-01 | 3.6777103E-02 | 1.8752161E-02 | 8.3852325E-03 | 2.7584346E-03 | 0.0000000E+00 |
| -7.000E-01 | 3.7663557E-02 | 1.9615661E-02 | 9.0668901E-03 | 3.0820460E-03 | 0.0000000E+00 |
| -6.000E-01 | 3.8550413E-02 | 2.0499507E-02 | 9.8305944E-03 | 3.4837469E-03 | 0.0000000E+00 |
| -5.000E-01 | 3.9437241E-02 | 2.1394545E-02 | 1.0679934E-02 | 3.9938569E-03 | 0.0000000E+00 |
| -4.000E-01 | 4.0323967E-02 | 2.2291735E-02 | 1.1607988E-02 | 4.6584232E-03 | 0.0000000E+00 |
| -3.000E-01 | 4.1210637E-02 | 2.3185446E-02 | 1.2586604E-02 | 5.5458368E-03 | 0.0000000E+00 |
| -2.000E-01 | 4.2097282E-02 | 2.4075568E-02 | 1.3559932E-02 | 6.7363759E-03 | 0.0000000E+00 |
| -1.000E-01 | 4.2983912E-02 | 2.4964268E-02 | 1.4484922E-02 | 8.1584797E-03 | 0.0000000E+00 |
| 0.000E+00 | 4.3870559E-02 | 2.5852988E-02 | 1.5390152E-02 | 9.2428635E-03 | 0.0000000E+00 |
| 0.000E+00 | 4.3870559E-02 | 2.5852988E-02 | 1.5390152E-02 | 9.2428635E-03 | 5.1143331E-03 |
| 1.000E-01 | 4.4757136E-02 | 2.6739735E-02 | 1.6283313E-02 | 1.0182052E-02 | 6.3812767E-03 |
| 2.000E-01 | 4.5643747E-02 | 2.7627407E-02 | 1.7179505E-02 | 1.1110058E-02 | 7.4202292E-03 |
| 3.000E-01 | 4.6530355E-02 | 2.8514896E-02 | 1.8073771E-02 | 1.2027346E-02 | 8.4040260E-03 |
| 4.000E-01 | 4.7416958E-02 | 2.9402242E-02 | 1.8966657E-02 | 1.2937963E-02 | 9.3605132E-03 |
| 5.000E-01 | 4.8303559E-02 | 3.0289473E-02 | 1.9858514E-02 | 1.3844054E-02 | 1.0300667E-02 |
| 6.000E-01 | 4.9190157E-02 | 3.1176611E-02 | 2.0749579E-02 | 1.4746881E-02 | 1.1230012E-02 |
| 7.000E-01 | 5.0076753E-02 | 3.2063674E-02 | 2.1640017E-02 | 1.5647247E-02 | 1.2151725E-02 |
| 8.000E-01 | 5.0963348E-02 | 3.2950674E-02 | 2.2529949E-02 | 1.6545696E-02 | 1.3067796E-02 |
| 9.000E-01 | 5.1849941E-02 | 3.3837622E-02 | 2.3419465E-02 | 1.7442613E-02 | 1.3979548E-02 |
| 1.000E+00 | 5.2736533E-02 | 3.4724526E-02 | 2.4308635E-02 | 1.8338283E-02 | 1.4887906E-02 |

Table 8c
Convergence of the $R_f + T_n$

| N | Original | W-e | Error Ratio |
|---|---|---|---|
| 22 | -6.485683389E-01 | -6.485683389E-01 | 1.000E+00 |
| 26 | 4.762675215E-01 | 5.579989726E-01 | 1.092E+00 |
| 30 | 6.075576519E-01 | 6.249103822E-01 | 2.018E+00 |
| 34 | 6.966171474E-01 | 3.953934070E-01 | 2.202E-01 |
| 38 | 7.662515839E-01 | 1.014371860E+00 | 1.489E-01 |
| 42 | 8.224944702E-01 | 1.042957034E+00 | 2.495E+00 |
| 46 | 8.689977656E-01 | 1.084299626E+00 | 1.404E+00 |
| 50 | 9.078613796E-01 | 1.074466852E+00 | 4.678E+00 |
| 54 | 9.391716470E-01 | 1.085391783E+00 | 3.312E+00 |
| 58 | 9.623331794E-01 | 1.044545990E+00 | 6.155E-01 |
| 62 | 9.778533603E-01 | 1.035787806E+00 | 1.877E+00 |
| 66 | 9.874321919E-01 | 1.924245741E+00 | 2.101E-02 |
| 70 | 9.930200889E-01 | 9.647870494E-01 | 5.658E-03 |
| 74 | 9.961698565E-01 | 1.002512093E+00 | 8.402E-02 |
| 78 | 9.979112922E-01 | 1.000064846E+00 | 7.131E-01 |
| 82 | 9.988642628E-01 | 1.000000759E+00 | 1.489E+01 |
| 86 | 9.993831144E-01 | 9.999995344E-01 | 4.241E+02 |
| 90 | 9.996649752E-01 | 9.999997913E-01 | 1.098E+03 |
| 94 | 9.998179838E-01 | 9.999999825E-01 | 8.005E+02 |
| 98 | 9.999010517E-01 | 9.999999958E-01 | 6.216E+03 |
| 102 | 9.999461709E-01 | 1.000000001E+00 | 9.628E+03 |
| 106 | 9.999706943E-01 | 1.000000001E+00 | 6.579E+04 |
| 110 | 9.999840334E-01 | 9.999999991E-01 | 7.592E+03 |
| 114 | 9.999912946E-01 | 1.000000001E+00 | 5.056E+03 |
| 118 | 9.999952502E-01 | 9.999999996E-01 | 3.980E+03 |
| 122 | 9.999974068E-01 | 9.999999999E-01 | 5.790E+03 |
| 126 | 9.999985833E-01 | 1.000000000E+00 | 1.379E+04 |
| 130 | 9.999992255E-01 | 1.000000000E+00 | 9.624E+03 |
| 134 | 9.999995764E-01 | 9.999999999E-01 | 1.051E+04 |
| 138 | 9.999997682E-01 | 9.999999999E-01 | 1.275E+04 |
| 142 | 9.999998731E-01 | 1.000000000E+00 | 2.465E+03 |
| 146 | 9.999999305E-01 | 1.000000000E+00 | 5.810E+04 |

## Appendix: Particular solution for the beam source

Here, we find the particular solution for the beam source by inserting

$$\tilde{q}_k(\tau) = -\left(\mathbf{T}^{-1}\mathbf{u}_0\right)_k e^{-\tau/\mu_0}$$

into the expression for the particular solution Eq(22) to give

$$\Theta_{Pk}(\tau) = W^{-1}(\lambda_k) \left[ \begin{array}{l} h(\lambda_k \tau) \int_{\tau}^{\tau_0} d\tau' h(\lambda_k(\tau_0 - \tau')) e^{-\tau'/\mu_0} + \\ + h(\lambda_k(\tau_0 - \tau)) \int_0^{\tau} d\tau' h(\lambda_k \tau') e^{-\tau'/\mu_0} \end{array} \right] \left(\mathbf{T}^{-1}\mathbf{u}_0\right)_k \quad \text{(A1)}$$

for the $k^{\text{th}}$ component of $\Theta_P(\tau)$. Expressing $h(\lambda_k \tau)$ and the Wronskian by Eqs(20b) and (20c) respectively, there results

$$\Theta_{Pk}(\tau) = \frac{1}{\lambda_k \sinh(\lambda_k \tau_0)} \left[ \begin{array}{l} \sinh(\lambda_k \tau) \int_{\tau}^{\tau_0} d\tau' \sinh(\lambda_k(\tau_0 - \tau')) e^{-\tau'/\mu_0} + \\ + \sinh(\lambda_k(\tau_0 - \tau)) \int_0^{\tau} d\tau' \sinh(\lambda_k \tau') e^{-\tau'/\mu_0} \end{array} \right] \left(\mathbf{T}^{-1}\mathbf{u}_0\right)_k .$$

(A2)

To analytically evaluate both integrals in this expression, consider the following integral:

$$J(\alpha; a, b) \equiv \int_a^b d\tau' \sinh(\lambda_k \tau') e^{\alpha \tau'}, \quad \text{(A3a)}$$

which evaluates to

$$J(\alpha;a,b) = \frac{1}{1-\frac{\lambda_k^2}{\alpha^2}}\left\{\left[\sinh(\lambda_k\tau') - \frac{\lambda_k}{\alpha}\cosh(\lambda_k\tau')\right]\frac{e^{\alpha\tau'}}{\alpha}\right\}_a^b. \qquad (A3b)$$

Introducing Eqs(A3b) into Eq(A2) and with subsequent simplification gives

$$\Theta_{Pk}(\tau) = \frac{\mu_0^2}{1-(\mu_0\lambda_k)^2}\frac{1}{\sinh(\lambda_k\tau_0)}\begin{bmatrix}-e^{-\tau/\mu_0}\begin{bmatrix}\sinh(\lambda_k\tau)\cosh(\lambda_k(\tau_0-\tau))+\\+\sinh(\lambda_k(\tau_0-\tau))\cosh(\lambda_k\tau)\end{bmatrix}+\\+e^{-\tau_0/\mu_0}\sinh(\lambda_k\tau)+\sinh(\lambda_k(\tau_0-\tau))\end{bmatrix}(T^{-1}u_0)_k$$

and from the identity for the sum of two arguments of the *sinh* function

$$\Theta_{Pk}(\tau) = -\frac{\mu_0^2}{1-\mu_0^2\lambda_k^2}\begin{bmatrix}e^{-\tau/\mu_0}-\\-e^{-\tau_0/\mu_0}\frac{\sinh(\lambda_k\tau)}{\sinh(\lambda_k\tau_0)}-\frac{\sinh(\lambda_k(\tau_0-\tau))}{\sinh(\lambda_k\tau_0)}\end{bmatrix}(T^{-1}u_0)_k. \qquad (A4)$$

Finally, by recognizing $h(\lambda_k\tau)$, the $k^{th}$ component is

$$\Theta_{Pk}(\tau) = -\frac{\mu_0^2}{1-\mu_0^2\lambda_k^2}\left[e^{-\tau/\mu_0}-e^{-\tau_0/\mu_0}h(\lambda_k\tau)-h(\lambda_k(\tau_0-\tau))\right](T^{-1}u_0)_k. \qquad (A5)$$

Before continuing, we must consider the possibility that $1/\mu_0 = \lambda_k$ for which the denominator in Eq(A5) vanishes. It can easily be shown that the numerator also vanishes giving an indeterminate form. Thus, using L'Hospital's rule, we find

$$\Theta_{Pk}(\tau) = \frac{\mu_0}{2}\left[\tau e^{-\tau/\mu_0}-\tau_0 e^{-\tau_0/\mu_0}h(\lambda_k\tau)\right](T^{-1}u_0)_k. \qquad (A6)$$

If for the moment we ignore this singular case, then the $\Theta_P(\tau)$ vector is

$$\Theta_P(\tau) = -diag\left\{\frac{\mu_0^2}{1-\mu_0^2\lambda_k^2}\left[e^{-\tau/\mu_0} - e^{-\tau_0/\mu_0}h(\lambda_k\tau) - h(\lambda_k(\tau_0-\tau))\right]\right\}T^{-1}u_0$$

from which the particular solution

$$\psi_P^+(\tau) = -Tdiag\left\{\frac{\mu_0^2}{1-\mu_0^2\lambda_k^2}\left[e^{-\tau/\mu_0} - e^{-\tau_0/\mu_0}h(\lambda_k\tau) - h(\lambda_k(\tau_0-\tau))\right]\right\}T^{-1}u_0$$

follows by Eq(23b). With some rearrangement, use of identities and identification of the *H*-matrix functions, there results

$$\psi_P^+(\tau) = -Tdiag\left\{\frac{\mu_0^2}{1-\mu_0^2\lambda_k^2}\right\}T^{-1}u_0 e^{-\tau/\mu_0} +$$

$$+ H(\tau)Tdiag\left\{\frac{\mu_0^2}{1-\mu_0^2\lambda_k^2}\right\}T^{-1}u_0 e^{-\tau_0/\mu_0} + \quad (A7)$$

$$+ H(\tau_0-\tau)Tdiag\left\{\frac{\mu_0^2}{1-\mu_0^2\lambda_k^2}\right\}T^{-1}u_0.$$

It is now convenient to let

$$R_P(\tau,\mu_0) \equiv -Tdiag\left\{\begin{array}{l}-\dfrac{\mu_0}{2}\tau;\ \lambda_k=1/\mu_0\\[2mm]\dfrac{\mu_0^2}{1-\mu_0^2\lambda_k^2};\ \lambda_k\neq 1/\mu_0\end{array}\right\}T^{-1}u_0 e^{-\tau/\mu_0}$$

including the singular case and the particular solution becomes

$$\psi_P^+(\tau) = R_P(\tau,\mu_0) - H(\tau)R_P(\tau_0,\mu_0) - H(\tau_0-\tau)R_P(0,\mu_0) \quad (A8)$$

as found above in Eq(51b) by simpler means.